\documentclass[superscriptaddress,nofootinbib]{revtex4-1}  

\usepackage{amsmath}
\usepackage{graphicx}
\usepackage{caption}
\usepackage{subcaption}
\usepackage{sidecap}
\usepackage{hyperref}
\usepackage{tabulary,booktabs}
\usepackage{enumitem,chngcntr}
\usepackage{color}

\newcommand{\ket}[1]{\mbox{$\left| #1 \right\rangle$}}

\begin{document}

\title{Analysis Tools for Next-Generation Hadron Spectroscopy Experiments}

\author{M.~Battaglieri} 
\affiliation{INFN Sezione di Genova, Via Dodecaneso 33, I--16146 Genova, Italy}

\author{B.~J.~Briscoe}
\affiliation{The George Washington University, 20052 Washington D.C., USA}

\author{A.~Celentano}
\affiliation{INFN Sezione di Genova, Via Dodecaneso 33, I--16146 Genova, Italy}

\author{S.-U.~Chung}
\affiliation{Technische Universit\"at M\"unchen, Physik Department, D--85748 Garching, Germany}
\affiliation{Department of Physics, Pusan National University, Busan 609-735, Republic of Korea}
\affiliation{Physics Department, Brookhaven National Laboratory, Upton, NY 11973, USA}

\author{A.~D'Angelo}
\affiliation{Universit\`a di Roma Tor Vergata and INFN, Sezione di Roma Tor Vergata, I--00133 Rome, Italy}

\author{R.~De Vita}
\affiliation{INFN Sezione di Genova, Via Dodecaneso 33, I--16146 Genova, Italy}

\author{M.~D\"oring}
\affiliation{The George Washington University, 20052 Washington D.C., USA}

\author{J.~Dudek}
\affiliation{Thomas Jefferson National Accelerator Facility, Newport News, VA 23606, USA}
\affiliation{Department of Physics, Old Dominion University, Norfolk, VA 23529, USA}

\author{S.~Eidelman}
\affiliation{Budker Institute of Nuclear Physics and
Novosibirsk State University, Novosibirsk, Russia}

\author{S.~Fegan}
\affiliation{INFN Sezione di Genova, Via Dodecaneso 33, I--16146 Genova, Italy}

\author{J.~Ferretti}
\affiliation{Dipartimento di Fisica and INFN, Universit\`a di Roma Sapienza, Piazzale A.~Moro 5, I--00185 Roma, Italy}

\author{A.~Filippi}
\affiliation{INFN Sezione di Torino, Via P. Giuria, 1, I-10125 Torino, Italy}

\author{G.~Fox}
\affiliation{School of Informatics and Computing, Indiana University, Bloomington, IN 47408, USA}

\author{G.~Galata}
\affiliation{INFN Sezione di Genova, Via Dodecaneso 33, I--16146 Genova, Italy}

\author{H.~Garc{\'{\i}}a-Tecocoatzi}
\affiliation{Instituto de Ciencias Nucleares, Universidad Nacional Aut\'onoma de M\'exico, 04510 M\'exico DF, M\'exico}
\affiliation{INFN Sezione di Genova, Via Dodecaneso 33, I--16146 Genova, Italy}

\author{D.~I.~Glazier}
\affiliation{SUPA, School of Physics and Astronomy, University of Glasgow, G12 8QQ, United Kingdom}

\author{B.~Grube}
\affiliation{Technische Universit\"at M\"unchen, Physik Department, D--85748 Garching, Germany}

\author{C.~Hanhart}
\affiliation{Institut f\"ur Kernphysik, Institute for Advanced Simulations, D--52425 J\"ulich, Germany}
\affiliation{ J\"ulich Center for Hadron Physics, Forschungszentrum J\"ulich, D--52425 J\"ulich, Germany}

\author{M.~Hoferichter}
\affiliation{Albert Einstein Center for Fundamental Physics, Institute for Theoretical Physics, University of Bern, Sidlerstrasse 5, CH--3012 Bern, Switzerland}
\affiliation{Institut f\"ur Kernphysik, Technische Universit\"at Darmstadt, D--64289 Darmstadt, Germany}
\affiliation{ExtreMe Matter Institute EMMI, GSI Helmholtzzentrum f\"ur Schwerionenforschung GmbH,D--64291 Darmstadt, Germany}

\author{S.~M.~Hughes}
\affiliation{Edinburgh University, Edinburgh EH9 3JZ, United Kingdom}

\author{D.~G.~Ireland}
\affiliation{SUPA, School of Physics and Astronomy, University of Glasgow, G12 8QQ, United Kingdom}

\author{B.~Ketzer}
\affiliation{Helmholtz-Institut f\"{u}r Strahlen- und Kernphysik, Rheinische Friedrich-Wilhelms-Universit\"{a}t Bonn, D--53115 Bonn, Germany}

\author{F.~J.~Klein}
\affiliation{The Catholic University of America, Washington, D.C. 20064}

\author{B.~Kubis}
\affiliation{Helmholtz-Institut f\"{u}r Strahlen- und Kernphysik, Rheinische Friedrich-Wilhelms-Universit\"{a}t Bonn, D--53115 Bonn, Germany}
\affiliation{Bethe Center for Theoretical Physics, Universit\"at Bonn, D--53115 Bonn, Germany}

\author{B.~Liu}
\affiliation{Institute of High Energy Physics, Beijing 100049, People's Republic of China}

\author{P.~Masjuan}
\affiliation{PRISMA Cluster of Excellence and Institut f\"ur Kernphysik, Johannes Gutenberg-Universit\"at Mainz, D-55099 Mainz, Germany}

\author{V.~Mathieu}\email{Corresponding author: mathieuv@indiana.edu}
\affiliation{Department of Physics, Indiana University, Bloomington, IN 47405, USA}
\affiliation{Center for Exploration of Energy and Matter, Indiana University, Bloomington, IN 47403, USA}

\author{B.~McKinnon}
\affiliation{SUPA, School of Physics and Astronomy, University of Glasgow, G12 8QQ, United Kingdom}

\author{R.~Mitchell}
\affiliation{Department of Physics, Indiana University, Bloomington, IN 47405, USA}

\author{F.~Nerling}
\affiliation{Helmholtz-Institut Mainz, GSI Helmholtzzentrum Darmstadt, Planckstrasse 1, 64291 Darmstadt}

\author{S.~Paul}
\affiliation{Technische Universit\"at M\"unchen, Physik Department, D--85748 Garching, Germany}

\author{J.~R.~Pel{\'a}ez}
\affiliation{Departamento de F\'{\i}sica Te\'orica II, Universidad Complutense, E--28040 Madrid, Spain}

\author{J.~Rademacker}
\affiliation{H H Wills Physics Lab, University of Bristol, Tyndall Avenue, Bristol BS8 1TL, United Kingdom}

\author{A.~Rizzo}
\affiliation{Universit\`a di Roma Tor Vergata and INFN, Sezione di Roma Tor Vergata, I--00133 Rome, Italy}

\author{C.~Salgado}
\affiliation{Norfolk State University, Norfolk, VA 23504}
\affiliation{Thomas Jefferson National Accelerator Facility, Newport News, VA 23606, USA}

\author{E.~Santopinto}
\affiliation{INFN Sezione di Genova, Via Dodecaneso 33, I--16146 Genova, Italy}

\author{A.~V.~Sarantsev}
\affiliation{Helmholtz-Institut f\"{u}r Strahlen- und Kernphysik, Rheinische Friedrich-Wilhelms-Universit\"{a}t Bonn, D--53115 Bonn, Germany}
\affiliation{Petersburg Nuclear Physics Institute, Gatchina 188300, Russia}

\author{T.~Sato}
\affiliation{Department of Physics, Osaka University, Toyonaka, Osaka 560-0043, Japan}

\author{T.~Schl\"uter}
\affiliation{Ludwig-Maximilians-Universit\"at, D--80799 M\"unchen, Germany}
\affiliation{Excellence Cluster Universe, Technische Universit\"at M\"unchen, D--85748 Garching, Germany}

\author{M.~L.~L.~da Silva}
\affiliation{Instituto de Fisica e Matematica, Universidade Federal de Pelotas, Caixa Postal 354, CEP 96010-090, Pelotas, RS, Brazil}

\author{I.~Stankovic}
\affiliation{Edinburgh University, Edinburgh EH9 3JZ, United Kingdom}

\author{I.~Strakovsky}
\affiliation{The George Washington University, 20052 Washington D.C., USA}

\author{A.~Szczepaniak}
\affiliation{Thomas Jefferson National Accelerator Facility, Newport News, VA 23606, USA}
\affiliation{Department of Physics, Indiana University, Bloomington, IN 47405, USA}
\affiliation{Center for Exploration of Energy and Matter, Indiana University, Bloomington, IN 47403, USA}

\author{A.~Vassallo}
\affiliation{INFN Sezione di Genova, Via Dodecaneso 33, I--16146 Genova, Italy}

\author{N.~K.~Walford}
\affiliation{The Catholic University of America, Washington, D.C. 20064}

\author{D.~P.~Watts}
\affiliation{Edinburgh University, Edinburgh EH9 3JZ, United Kingdom}

\author{L.~Zana}
\affiliation{Edinburgh University, Edinburgh EH9 3JZ, United Kingdom}

\maketitle

\tableofcontents

\clearpage
\section*{Preface}

The series of workshops on {\it New Partial-Wave Analysis Tools for
  Next-Generation Hadron Spectroscopy Experiments} was initiated with
the ATHOS~2012 meeting, which took place in Camogli, Italy, June 20--22,
2012.  It was followed by ATHOS~2013 in Kloster Seeon near Munich,
Germany, May 21--24, 2013. The third, ATHOS3, meeting is planned for April 13--17, 2015 
at The George Washington University Virginia Science and Technology Campus, USA.

The workshops focus on the development of amplitude analysis tools for
meson and baryon spectroscopy, and complement other programs in hadron
spectroscopy organized in the recent past including the INT-JLab
Workshop on Hadron Spectroscopy in Seattle in 2009, the International
Workshop on Amplitude Analysis in Hadron Spectroscopy at the
ECT*-Trento in 2011, the School on Amplitude Analysis in Modern
Physics in Bad Honnef in 2011, the Jefferson Lab Advanced Study
Institute Summer School in 2012, and the School on Concepts of Modern
Amplitude Analysis Techniques in Flecken-Zechlin near Berlin in
September 2013.

The aim of this document is to summarize the discussions that took
place at the ATHOS~2012 and ATHOS~2013 meetings.  We do not attempt a
comprehensive review of the field of amplitude analysis, but offer a
collection of thoughts that we hope may lay the ground for such a
document.

\vfill

\noindent The Editorial Board \vskip 10pt

\noindent Marco Battaglieri

\noindent Bill J.~Briscoe

\noindent Su-Urk Chung

\noindent Michael D\"oring 

\noindent J\'ozef Dudek 

\noindent Geoffrey Fox

\noindent Christoph Hanhart

\noindent Martin Hoferichter

\noindent David G.~Ireland 

\noindent Bernhard Ketzer

\noindent Bastian Kubis

\noindent Vincent Mathieu

\noindent Ryan Mitchell 

\noindent Jos\'e R.\ Pel{\'a}ez 

\noindent Elena Santopinto

\noindent Adam Szczepaniak

\section{Introduction}
\vskip-5pt [C.~Hanhart, M.~Hoferichter, B.~Kubis] \vskip10pt

Quantum Chromodynamics (QCD), the fundamental theory of the strong
interactions, defines the interactions of quarks and gluons, both
types carrying the so-called color charge, which form the fundamental
constituents of hadrons.\footnote{All composite objects of quarks and
  gluons that are therefore subject to the strong interaction with no
  net color charge are called hadrons.} At high energies, these
partons become asymptotically free, and systematic calculations based
on perturbation theory in the strong coupling constant are possible
and extremely successful. However, especially inside light hadrons
that are in the focus of this manuscript, the average energies and
momenta of partons are below the scale at which perturbation theory
can be justified, and hadron properties are determined by
interactions 
that are genuinely non-perturbative in nature. In particular, the bulk
of hadron masses originates from gluonic self-interactions, which lead
to forces that bind the constituents within distances smaller than
$10^{-15}\,\text{m}$ in a way that only allows objects neutral with
respect to the color charge to exist as physical, asymptotic
states---a phenomenon known as confinement.  As a consequence, the
elementary degrees of freedom of the underlying theory only manifest
themselves indirectly in the physical spectrum, which instead is built
from composite, colorless hadrons.  Just as atomic spectroscopy was
instrumental in elucidating the underlying electromagnetic
interactions, hadron spectroscopy is therefore the foremost laboratory
for studying the implications of QCD.
   
While for many years the quark model has provided the main template
for the spectrum of hadrons, recent developments in lattice
simulations on the one side and effective-field-theory methods on the
other have opened new avenues for investigations of hadron properties
that are rooted in QCD.  One of the most mysterious parts of the
spectrum concerns the phenomenology of low-energy gluons and thus a
complete mapping of gluonic excitations---that may manifest themselves
either in hybrid states (states with both quarks and gluons as active,
valence degrees of freedom) or in glueballs (states formed from
gluons only)---is a central part of the present and future
investigations of the hadron spectrum.

The anticipated accuracy of the next-generation hadron spectroscopy
experiments will in principle allow for the identification of hadronic
resonances for which either a reliable determination of their
resonance parameters has proven elusive or even their very existence
could not be unambiguously established before. Frequently, their
identification is complicated by the occurrence of overlapping
resonances, pole positions far in the complex plane, or weak couplings
to the channels experimentally accessible.  The main challenges
include the development of parameterizations and their incorporation
into partial-waves analyses that respect the theoretical constraints
and allow for a reaction-independent determination of pole positions
and residues, which uniquely characterize the properties of a given
resonance.  In this document we review some aspects of the theoretical
and phenomenological underpinning of experimental data analyses that
aim at extracting hadron resonance parameters in a controlled way.

Beyond providing a deeper understanding of the inner workings of QCD,
a theoretical control over hadronic final-state interactions is also
essential to employ the decays of heavy mesons for the hunt of physics
beyond the Standard Model of particle physics (SM), which are driven
by the electroweak interactions: in order to explain the
matter--antimatter asymmetry of the universe, an amount of CP~violation 
is necessary that exceeds that of the SM by many orders of
magnitude. Thus, additional CP~violation has to be present, and it has
to exceed the SM predictions dramatically.

If present, CP~violation in the decay of heavy mesons will show up as
a complex phase, and therefore relies on interference of different
amplitudes.  As the observation of CP~asymmetries in (partial) decay
rates depends on both weak and strong phase differences, a more
accurate understanding of the latter necessarily leads to an improved
determination of the former, and resonating strong final states
provide ideal enhancement factors for (probably very small) weak
asymmetries.  Therefore, the decay of a heavy meson into three or more
light mesons appears to provide an ideal environment for CP~studies
due to the presence of a large number of meson resonances in the phase
space available.  Furthermore, besides enhancing the CP~signals, the
non-trivial distribution of the strong phase motion over the Dalitz
plot allows for a test of systematics, and provides some sensitivity
to the operator structure of the CP-violating source underlying the
transition.

This twofold perspective of amplitude analyses should be kept in mind
throughout this document: while a strong motivation clearly consists
in understanding the spectrum of QCD as such, there is a strong
benefit from making the results available for communities more
concerned with the investigation of electroweak interactions and New
Physics searches in hadronic environments.

\subsection{Quark model}
\label{sec:QM}
\vskip-5pt [V.~Mathieu] \vskip10pt
 
The Quark Model was originally introduced as a classification scheme
to organize the hadron spectrum. Since its introduction, significant
progress has been made in the understanding of QCD, and while there is
no formal relation between constituent quarks and the QCD degrees of
freedom, the lattice QCD hadron spectrum closely resembles that of the
quark model.  In the quark model, mesons are bound states of a
valence, constituent quark and antiquark, while baryons contain three
quarks.  Quantum numbers of quark model bound states are obtained by
combining the quantum numbers of the individual quark constituents,
e.g.\ their spins and angular momenta. For example, the meson
spin $J$ is given by the vector sum of quark--antiquark spin $s$ and
orbital angular momentum $l$. Meson parity $P$ and, for neutral
states, charge conjugation $C$ are given by $P=(-1)^{l+1}$,
$C=(-1)^{l+s}$, respectively. It thus follows that certain
combinations of total spin $J^{PC}$, $0^{--}$, $0^{+-}$, $1^{-+}$,
$2^{+-}$, $3^{-+}$, $\ldots$, do not correspond to a quark--antiquark
pair.  These are referred to as \textit{exotic}.  There are no exotic
baryons in a corresponding sense, i.e.\ three quarks can be
combined to give any combination of a half-integer spin and parity.
In addition, taking into account quark flavors the quark model
arranges hadrons into flavor multiplets with mass degeneracies broken
by the quark masses.

The classification of the well-established light mesons according to
the quark model is summarized in Table~\ref{tab:PDG} taken from the
Review of Particle Physics~\cite{Agashe:2014kda}.  Indeed, most of
the observed resonances fit into the quark-model pattern, although
several states including the $\rho_2$ or the $b_3$ are missing.  There
are also well-established resonances that do not fit the quark-model
classification. These include for example states with
$J^{PC}=0^{++}$ quantum numbers, such as the $f_0(500)$.
    
Hadron resonances can also be classified by the Regge trajectories
they belong to. For example, for mesons, Regge trajectories are
labeled by signature $\tau=(-1)^J$, naturality $\eta=P(-1)^J$, and
also by isospin $I$ and $G$-parity $G=C(-1)^I$.

\begin{table}
  \begin{center}
    \renewcommand{\arraystretch}{1.3}
    \begin{tabular}{ccccccc}
      \toprule
\hline
      $n ^{2s+1}\ell_J$ & $J^{PC}$  & $I=1$ & $I=1/2$ &$I=0$&  $I=0$  \\
      \midrule
\hline
      $1 ^1S_0$ & $0^{-+}$ & $\pi$ & $K$ & $\eta$ & $\eta^\prime$ \\
      $1 ^3S_0$ & $1^{--}$ & $\rho(770)$ & $K^*(892)$ & $\omega(782)$ & $\phi(1020)$ \\
      \midrule
\hline
      $1 ^1P_1$ & $1^{+-}$ & $b_1(1235)$ & $K_1(1400)$ & $h_1(1170)$ & $h_1(1380)$ \\
      $1 ^3P_0$ & $0^{++}$ & $a_0(1450)$ & $K_0^*(1430)$ & $f_0(1370)$ & $f_0(1710)$ \\
      $1 ^3P_1$ & $1^{++}$ & $a_1(1260)$ & $K_1(1270)$ & $f_1(1285)$ & $f_1(1420)$\\
      $1 ^3P_2$ & $2^{++}$ & $a_2(1320)$ & $K_2^{*}(1430)$ & $f_2(1270)$ & $f_2^\prime(1525)$ \\
      \midrule
\hline
      $1 ^1D_2$ & $2^{-+}$ & $\pi_2(1670)$ & $K_2(1770)$ & $\eta_2(1645)$ & $\eta_2(1870)$ \\
      $1 ^3D_1$ & $1^{--}$ & $\rho(1700)$ & $K^*(1680)$ & $\omega(1650)$ & \\
      $1 ^3D_2$ & $2^{--}$ &  & $K_2(1820)$ &  &  \\
      $1 ^3D_3$ & $3^{--}$ & $\rho_3(1690)$ & $K_3^*(1780)$ & $\omega_3(1670)$ & $\phi_3(1850)$\\
      \midrule
\hline
      $1 ^1F_3$ & $3^{+-}$ &  &  & &  \\
      $1 ^3F_2$ & $2^{++}$ &  & $K_2^*(1980)$ & $f_2(1910)$ & $f_2(2010)$ \\
      $1 ^3F_3$ & $3^{++}$ &  & $K_3(2320)$ & &  \\
      $1 ^3F_4$ & $4^{++}$ & $a_4(2040)$ & $K_4^{*}(2045)$ & & $f_4(2050)$ \\
      \bottomrule
\hline
    \end{tabular}
\caption{Well-established mesons classified according to the quark model. \label{tab:PDG}}
\end{center}
\end{table}

\subsection{Lattice QCD and the hadron spectrum}
\vskip-5pt [J.~Dudek, M.~D\"oring] \vskip10pt

Lattice QCD is a first principles numerical approach to QCD which
considers the field theory evaluated on a finite grid of
points. Supercomputers are used to Monte-Carlo sample a finite, but
large, number of gluon field configurations according to their
importance in the QCD Euclidean path integral. Color-singlet
correlation functions can then be computed using this ensemble of
configurations, with the mean and variance over the ensemble providing
an estimate and an uncertainty. The discrete spectrum of eigenstates
of the theory can be extracted from the time-dependence of correlation
functions.

In principle this is a systematically improvable approach to
QCD. Calculations can be performed for a range of lattice spacings,
$a$, and an extrapolation $a \to 0$ performed. Similarly the behavior
with increasing finite volume can be studied. In practice, the low
mass of the physical $u$ and $d$ quarks provides a challenge---the
numerical algorithms used to generate gluon field configurations and
to compute quark propagation scale badly with decreasing quark
mass. Furthermore, since very light quarks imply very light pions with
large Compton wavelengths, there is a need to increase the size of the
lattice volume as the quark mass decreases. For fixed lattice spacing
this requires more points in the grid and thus increased computation
time.

\begin{figure}
\centering
  \includegraphics[width=0.55\linewidth]{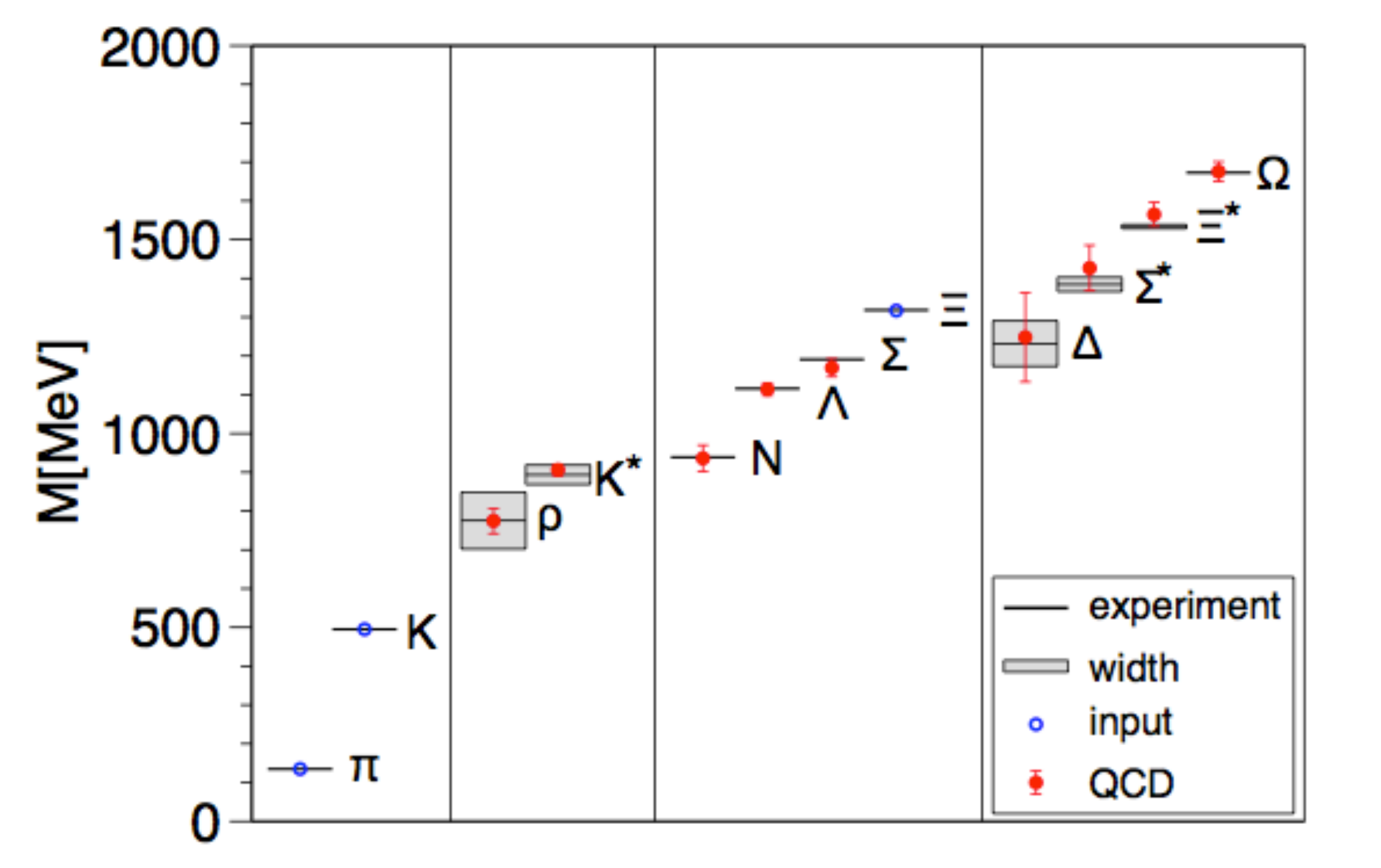}
  \caption{The light hadron spectrum of QCD computed using lattice
    techniques in~\cite{Durr:2008zz}. \label{fig:SpectrumBMW} }
\end{figure}

\begin{figure}
  \centering
  \includegraphics[width=\textwidth]{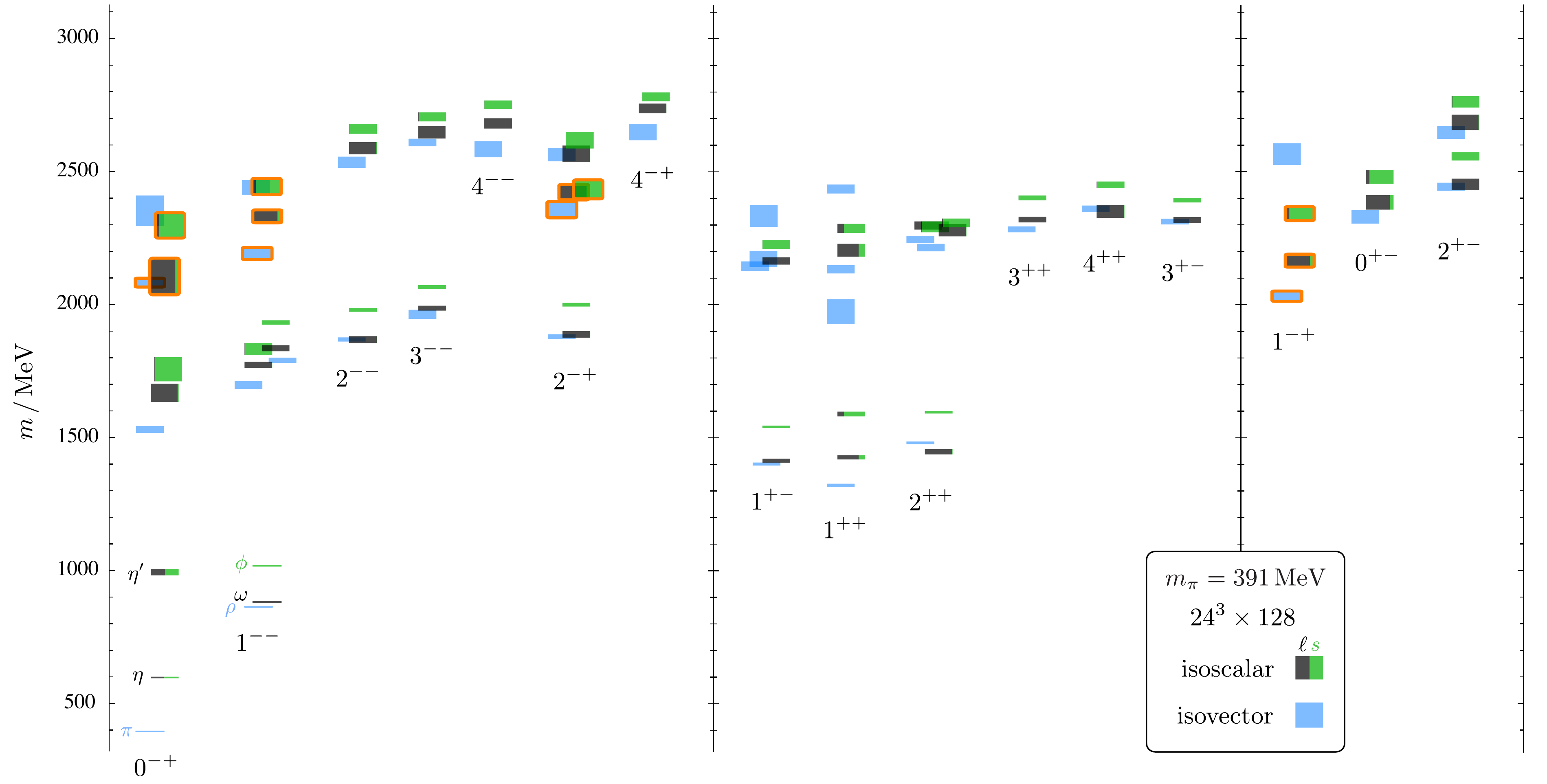}
\caption{\label{fig:lqcd_mesons}
  Isoscalar and isovector meson spectrum determined in a lattice QCD
  calculation with $m_\pi = 391\,\mathrm{MeV}$~\cite{Dudek:2013yja}.}
\end{figure}

For relatively simple quantities like the masses of the lightest
stable hadrons, precision calculations considering all the above
systematic variations have recently been carried out. An example is
presented in Fig.~\ref{fig:SpectrumBMW}. In the case of {\it excited}
hadrons, the state of the art is not yet at this level, with
calculations typically being performed at a single (albeit small)
lattice spacing, and with light quark masses chosen to be somewhat
above the physical value. Fig.~\ref{fig:lqcd_mesons} presents an
example of recent progress in determining the excited isoscalar and
isovector meson spectrum. This calculation has approximately physical
strange quarks but light quarks somewhat heavier than physical such
that the pion has a mass of 391\,MeV~\cite{Dudek:2013yja,
  Dudek:2010wm,Dudek:2011tt}.

Fig.~\ref{fig:lqcd_mesons} shows a detailed spectrum of excited states
of various $J^{PC}$, with many of the observed experimental
systematics being reproduced, as well as those of the
$n\,^{2S+1}\!L_J$ $q\bar{q}$ quark model. A clear set of exotic
$J^{PC}$ states are extracted with the isovector spectrum featuring a
lightest $1^{-+}$ roughly 1.3\,GeV heavier than the $\rho$
meson. Slightly heavier than the $1^{-+}$ is a single $0^{+-}$ state
and two $2^{+-}$ states, and these observations have been shown to be
robust with increasing quark mass. Examination of the type of
quark--gluon operator constructions which have large overlap with these
exotic states suggests that they are hybrid mesons with $q\bar{q}$ in
a color octet coupled to a chromomagnetic gluonic excitation. Such a
construction can also generate non-exotic hybrid mesons, and indeed
such states with $J^{PC}=0^{-+}$, $2^{-+}$, and $1^{--}$ are
identified in the calculation (highlighted in orange in
Fig.~\ref{fig:lqcd_mesons}). Calculation in the charmonium
sector~\cite{Liu:2012ze, Guo:2008yz} shows similar conventional meson and hybrid-meson systematics.

The baryon spectrum has been computed using related
techniques~\cite{Edwards:2011jj, Edwards:2012fx}, see also 
Refs.~\cite{Bulava:2009jb, Alexandrou:2013fsu, Engel:2013ig, Mahbub:2012ri}. Hybrid baryons,
which cannot have exotic quantum numbers, have been
predicted~\cite{Dudek:2012ag} with a quantum number distribution and
operator overlaps that suggest the same chromomagnetic gluonic
excitation is at work.

Computing the spectrum of glueballs is relatively straightforward
within the pure-glue theory where the existence of quarks is
ignored. Glueball operators can be constructed out of gluon fields and
the spectrum extracted from correlation functions. The spectra so
determined in~\cite{Morningstar:1999rf, Chen:2005mg} show that the
lightest glueballs have non-exotic $J^{PC}$ with a lightest $0^{++}$
and somewhat heavier a $2^{++}$ and a $0^{-+}$. However in QCD, with
quarks, these glueball basis states should appear embedded within a
spectrum of isoscalar mesons, possibly strongly mixed with $q\bar{q}$
basis states. Such calculations have proven to be very challenging,
for example the calculation in~\cite{Dudek:2013yja} was not able to
observe any states having strong overlap with glueball operators,
which produced statistically noisy correlation functions. In short the
role of glueballs in the meson spectrum has not been determined in
lattice QCD.

Returning to Fig.~\ref{fig:lqcd_mesons}, although a lot of the correct
physics is present, including annihilation of $q\bar{q}$ pairs and the
corresponding mixing of hidden-light and hidden-strange
configurations, the calculations are clearly not complete. Most of the
states extracted should in fact be unstable resonances decaying into
multi-meson final states. In fact, within a finite-volume theory,
there cannot be a continuum of multi-meson states, rather there must be
a discrete spectrum and the volume-dependence of this spectrum can be
mapped onto hadron scattering amplitudes~\cite{Luscher:1990ux,
  Rummukainen:1995vs, Kim:2005gf, Christ:2005gi,Feng:2010es,Lang:2011mn, 
  Doring:2011vk, Doring:2012eu, Gockeler:2012yj, 
  Dudek:2014qha}---often
  calling for the inclusion of inelasticities~\cite{Doring:2011vk,Lage:2009zv,
  Bernard:2010fp,Briceno:2012yi,Hansen:2012tf, Guo:2012hv}. 
  The full richness of this spectrum was
not resolved in~\cite{Dudek:2013yja} as only quasi-local
$q\bar{q}$-like operator constructions were used, and these have very
poor overlap onto multi-meson states.

\begin{figure}
\centering
  \includegraphics[width=0.55\linewidth]{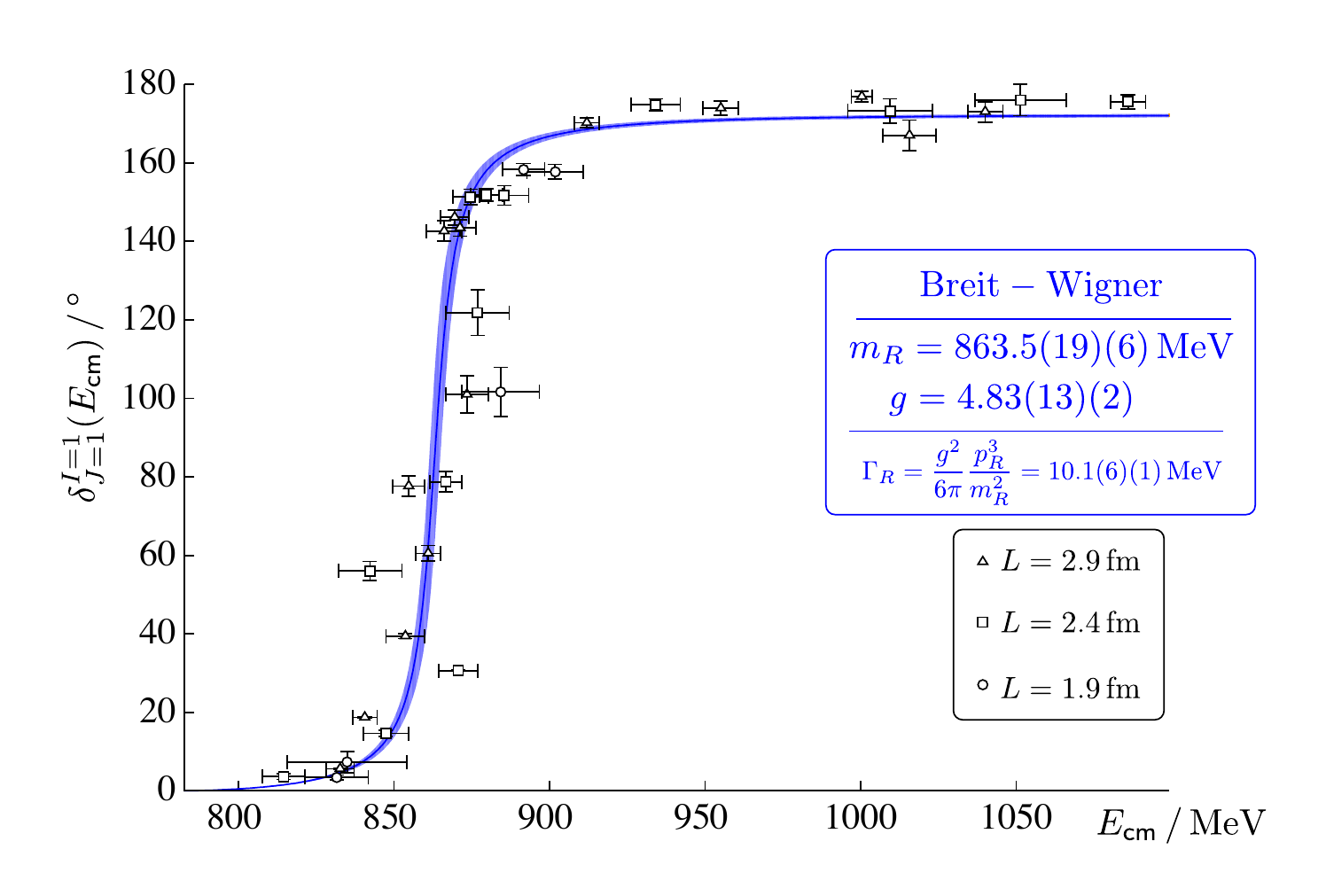}
  \caption{The isospin $1$, $P$-wave $\pi\pi$ scattering phase shift
    determined from the discrete spectrum in three different lattice
    volumes. Calculation performed with quark masses such that $m_\pi
    = 391\,\mathrm{MeV}$~\cite{Dudek:2012xn}. \label{fig:rhopipi} }
\end{figure}

The current frontier in lattice QCD calculations of hadron
spectroscopy involves the inclusion of operators that efficiently
interpolate multi-meson states and the extraction of the complete
discrete spectrum of states in a finite volume. An example of what can
currently be achieved is presented in Fig.~\ref{fig:rhopipi}. By computing
the complete low-energy spectrum of states with isospin~$1$ in
multiple finite volumes, and applying the finite-volume
formalism~\cite{Luscher:1990ux, Rummukainen:1995vs} to determine the
elastic $P$-wave scattering phase shift, a rapid rise characteristic
of a resonance can be observed. Fitting the phase shift with a simple
Breit--Wigner form yields an estimate of the $\rho$ resonance mass and
width in a version of QCD where the pion mass is 391\,MeV~\cite{Dudek:2012xn}. 
See Refs.~\cite{Feng:2010es, Lang:2011mn, Pelissier:2012pi, Aoki:2011yj, Frison:2010ws, Gockeler:2008kc} 
for other studies on the 
$\rho$ decay using the L\"uscher formalism. 
Ongoing calculations are addressing higher
resonances which can decay into multiple channels. The first 
coupled-channel lattice QCD calculation has been completed 
recently~\cite{Dudek:2014qha}. Concepts are developed 
to deal with three-body and
higher scattering in the finite volume~\cite{Polejaeva:2012ut,Briceno:2012rv, 
Kreuzer:2012sr, Hansen:2013dla}.

In the near future we envisage the possibility of using the same
scattering amplitude parameterizations to describe experimental data
and the finite-volume spectra of QCD computed using lattice
techniques.

\section{Experiments}
\label{ch:experiments} 

\vskip-5pt [M.~Battaglieri, D.~Ireland, B.~Ketzer, R.~Mitchell, A.~V.~Sarantsev] \vskip10pt

A number of hadron spectroscopy experiments are currently in
operation, and several new ones are expected to come online in the
near future. At low energies, fixed-target experiments studying elastic or
quasi-elastic meson--nucleon scattering  
reactions have been instrumental for baryon spectroscopy. Most of our
information about $N^*$ and $\Delta$ resonances stems from direct
production in elastic and inelastic $\pi N$ scattering experiments
from more than 30 years ago. In these reactions, also referred to as
$s$-channel production or formation, the beam and the target merge to
produce the 
resonance, which then subsequently decays. Phase-shift analysis in the
elastic region is a well-defined procedure that yields the scattering
amplitude from the experimental data with only a few discrete
alternative solutions. Elastic pion--nucleon scattering still 
provides the foundation 
for precise partial-wave analyses of baryon resonances, but an
improvement to the data set of meson-induced reactions is only likely
to be possible with the use of secondary hadron beams, produced at
future facilities such as J-PARC or an electron--ion collider (EIC). 

Current fixed-target experiments at  
electron machines like CLAS (JLAB), A2 (MAMI), or CBELSA/TAPS (ELSA)
mainly use photoproduction of 
resonances on proton targets to study baryon excitations.  
The JLab12 
upgrade, with the two new detectors GlueX and CLAS12, will enable a
dedicated program of spectroscopy in which a major goal will be the
discovery of hybrid mesons and baryons containing light quarks. 

High-energy fixed-target experiments with hadron beams like VES (IHEP
Protvino) or COMPASS (CERN) study $t$-channel reactions of the beam
particles with protons or nuclear targets. The target particle merely
serves as a strong-interaction partner and takes up the
recoil. 

Experiments with hadrons containing charm or bottom quarks require higher
center-of-mass energies, and are performed either at $e^+e^-$
colliders, such as BES-III (BEPC), KEDR (VEPP-4M), and Belle-II (KEKB), 
at $pp$ colliders such as ALICE, ATLAS, CMS, and LHCb (LHC), or
using $p\overline{p}$ annihilations like at the future PANDA
experiment at the High-Energy Storage Ring (HESR) for antiprotons at
FAIR. 

Key ingredients for next-generation experiments in hadron spectroscopy
are:
\begin{itemize}
\item detectors for both charged and neutral particles, with
  excellent resolution and particle identification capability;
\item beam energies high enough to ensure sufficient phase space for
  production;
\item high statistics, for sensitivity to production cross sections at
  picobarn level;
\item networks of experimentalists and theorists working together for
  the development of common analysis tools.
\end{itemize}

\subsection{Fixed-target experiments}
\indent

Fixed-target experiments with primary electron or proton beams as well
as secondary
meson ($\pi$, $K$) or photon 
beams have been at the heart of mapping out and understanding the
light hadron spectrum for more than 30 years. 

Depending on the incident
particle, the energy and  the final-state kinematics, different
mechanisms contribute to the production of excited meson states:
$s$-channel resonance formation at low energies, and production
reactions involving a recoil particle 
at higher energies. At very high energies, $t$-channel reactions like 
diffraction,  
central production,  
or photoproduction involving quasireal photons dominate. 

\subsubsection{Hadron beams: COMPASS, VES, and PANDA}
\indent

COMPASS~\cite{Abbon:2007pq, Abbon:2014aex} is a high-energy hadron physics experiment
at the Super Proton 
Synchrotron at CERN involving about 220 physicists from 13 countries
and 24 
institutions. One of the purposes of this experiment is to study
hadron spectroscopy using high-intensity hadron beams of
$150$--$250\,\mathrm{GeV}$ by diffractive, central, and Coulomb
production reactions. Final states containing charged and neutral
particles are detected with high resolution over a wide angular range,
provided by a two-stage magnetic spectrometer equipped with precision
vertex detectors, 
charged-particle tracking, particle identification, and calorimetry. A
uniform acceptance for both charged and neutral particles as realized
in COMPASS is mandatory for a reliable partial-wave analysis. 

One of the goals of COMPASS is to understand and map out the spectrum
of mesons up to masses of about $2.5\,\mathrm{GeV}$ with high statistical
accuracy, and to look for possible signatures of states which cannot
be explained within the constituent quark model, e.g.\ multi-quark and
hybrid states or glueballs. The non-$q\overline{q}'$ nature of a
resonance may be identified either through exotic quantum numbers,
which require additional contributions beyond a quark and an
antiquark, or via an overpopulation of states compared to expectations
from the quark model. The latter approach, however, requires the
unambiguous identification of all states of a given $J^{PC}$ nonet, a
task which has been achieved so far only for the $L=0,1$ mesons, cf.\ Table~\ref{tab:PDG}. 

In a first analysis of the $\pi^-\pi^-\pi^+$ final state from the 
scattering of  $190\,\mathrm{GeV}$ $\pi^-$ on a Pb target, recorded in 2004,  
a clear signal in intensity and phase motion in the
$J^{PC}=1^{-+}$ $I^G=1^+$ $\rho\pi$ $P$~partial wave has been
observed by COMPASS~\cite{Alekseev:2009xt}, consistent with the
$\pi_1(1600)$.  
However, a large background, possibly due to Deck-like processes, is also
present in the data, and will have to be properly taken into account
in a more refined partial-wave analysis.  
Two orders of magnitude more data with
pion and proton beams on proton and nuclear targets have been
collected in 2008, 
2009, and 2012, recording samples of various final
states: 54M events for $3\pi$, 144k events for $K\bar K\pi\pi$, 116k
events for $\eta\pi$, 39k events for $\eta'\pi$, etc. 
A signal for an exotic $1^{-+}$ state is also observed in the
$\eta'\pi$ final state~\cite{Adolph:2014uba}. As for the $3\pi$ final state, however, the
clear distinction between resonant and nonresonant contributions
requires a more reliable model for the background processes to be
included in the fit to the spin-density matrix. 
The new data for the first time allow an analysis in narrow bins of
invariant mass and 4-momentum transfer $t$, which has the power to shed more
light on the relative contribution of resonant and 
nonresonant processes in this and other waves, e.g.\ the well-known
$a_1(1260)$. 

The VES experiment uses a $28\,\mathrm{GeV}$ secondary pion beam from the U-70
proton synchrotron at IHEP
Protvino incident on a
Be target to study light-meson resonances decaying to neutral and
charged pions. The analysis techniques used are very similar to the
ones employed at COMPASS. 

PANDA is one of the major projects planned for the
FAIR-Facility in Darmstadt. FAIR is an extension of the existing Heavy
Ion Research Lab (GSI) and is expected to start operation in
2018. PANDA studies interactions between antiprotons and fixed target
protons and nuclei in the momentum range of $1.5$--$15$\,GeV using the
high-energy storage ring HESR. The PANDA collaboration, with more than
450 scientists from 17 European countries, intends to do basic
research on various topics around the weak and strong forces, exotic
states of matter, and the structure of hadrons.

\subsubsection{Electron beams: CLAS, ELSA, MAMI, SPring-8, and JLAB12}  
\indent

In the last 20 years electron accelerators such as CEBAF at JLab, ELSA at
Bonn, MAMI at Mainz, and SPring-8 in Japan, have considerably improved
in the delivery of electron and photon beams of high intensity and  quality to
enable coincidence measurements for hadron spectroscopy. New detectors
and targets  have been designed and
commissioned. We are now in a situation where the photo- and
electroproduction of pseudoscalar mesons carry the highest potential
to investigate the baryonic spectrum. In addition to the resonance
positions and strong residues, which describe couplings to decay
channels, the electromagnetic couplings and transition form factors
are also being investigated.

Pseudoscalar  (e.g.~$\pi$, $\eta$,
$K$, and $\eta^\prime$) photoproduction is one of the cleanest 
ways to study direct baryon production. In fact,
this reaction is described by a set of only four
transition amplitudes~\cite{Chew:1957tf} (invariant, spin,
transversity, or helicity amplitudes). With the developments in beams
and target mentioned above, we are now very near to being able to
extract these amplitudes (up to an overall phase) from a combination
of polarization experiments.

It is also very important that data obtained with a proton target are
complemented with data from the neutron, albeit in quasi-free
production from a nuclear target. At JLab, data taken with deuterium
targets, in particular the HD-ICE target, will provide information on
the $\gamma$-neutron couplings of excited states by extracting single-
and double-polarization observables. Complementary campaigns at ELSA
and MAMI will also provide important data sets.

Data obtained from the neutron bound in the nuclear targets (mostly
deuteron) will be sensitive to nuclear effects such as Fermi motion
and final-state scattering. Great care needs to be taken in unfolding
the desired amplitudes and multipoles. At present the experimental
information from proton reactions is substantially larger then that
from (quasi-)neutron reactions (the $\gamma N \to \pi N$ database is
just 15\% of the proton database), and this difference is especially
acute for polarized experiments. Only with sufficient data on both
proton and neutron targets, can one hope to disentangle the isoscalar
and isovector electromagnetic couplings of the various $N^*$ and
$\Delta^*$ resonances, as well as the isospin properties of the
non-resonant background amplitudes.

The search for mesons with exotic quantum numbers is the primary aim
of the GlueX experiment at a future 12\,GeV upgrade of Jefferson
Laboratory. The GlueX experiment will map out the meson spectrum with
unprecedented statistics using photoproduction, a complementary
reaction mechanism to others studied so far (which include
hadroproduction with pion, kaon, or proton beams, or heavy meson
decays). With 9\,GeV photons the mass range extends up to 2.5--3\,GeV
and will cover the region where the light exotic multiplet is
expected. A complementary meson spectroscopy program will be carried out
at Hall-B with the new CLAS12 detector. The technique,
electroproduction at very low $Q^2$ (0.01--0.1\,GeV$^2$), provides a
high photon flux as well as a high degree of linear polarization and
represents a competitive and complementary way to study the meson
spectrum and production mechanisms with respect to real
photoproduction experiments. After a calibration period, the detector
will begin to record data in 2015/16. Both GlueX and CLAS12 physics
programs will start in conjunction with the analysis of the golden
channels $\eta\pi$, $\eta^\prime\pi$, and $3\pi$ for the detection of
hybrid mesons. A detailed theoretical study on these channels is then
required in the near future for the success of these experiments.

\subsection{Annihilation reactions: Belle-II, BES-III, CMD-3, LHCb, and SND}
\indent

Annihilation of $e^+ e^-$ and $p \bar p$ have been historically
important additions to the host of reactions in hadron
spectroscopy. The early experiments in the SLAC-LBL $e^+ e^-$ storage
ring~(SPEAR) produced many of the first measurements in the charmonium
spectrum. They were followed by, among others, CLEO, BaBar, Belle,
BES-III, CMD-3 and SND, with the latter three still in operation. Charmonium
decay data sets have been supplemented by bottomonium decay data and
open-flavor $D$ and $B$~meson decays. Proton--antiproton annihilation
was studied at the Low Energy Antiproton Ring~(LEAR) at CERN and new
experiments at center-of-mass energies above charm threshold are planned
for the FAIR facility (see the description of PANDA in the fixed target experiments section). LHCb is exploiting the highest energy ever reached by the LHC
to produce a huge number of mesons and study their decays.

There has recently been a dramatic renewal of interest in the subjects
studied by these experiments. This renaissance has been driven in part
by experimental reports of $D^0\bar D^0$ mixing and the discovery of
narrow $D_{sJ}$ states and a plethora of charmonium-like $XYZ$ states
at the $B$ factories, as well as the observation of an intriguing
proton--antiproton threshold enhancement and the possibly-related
$X(1835)$ meson state at BES-II.  Many of these studies have relied on
amplitude analysis techniques and phenomenology.  For example, during
the $B$~factory age, the program to extract weak interaction parameters
(such as the CKM matrix elements) or to study New Physics effects went
through the analysis of decays with final states with at least three
particles. Light hadron final-state interactions bring in phases,
which interfere with the weak phases and have to be included in an
amplitude analysis.  The $D^0 \to K_s \pi\pi$ amplitude, as one
example, depends on the weak CKM phase $\gamma$, which can only be
extracted if the strong $K\pi$ and $\pi\pi$ phases are
known~\cite{Gerard:1997kv,Gerard:1998un}. 

Here we briefly describe five facilities that are currently in operation or
are planned. 
The BES-III experiment at BEPCII in Beijing~\cite{Asner:2008nq}, which started operation in the summer 2008, has accumulated data samples corresponding to 1.3 billion $J/\psi$ decays, 0.6 billion $\psi(3686)$ decays, 2.9 fb$^{-1}$ at the peak of the $\psi(3770)$ resonance, and around 4\,fb$^{-1}$ above 4\,GeV.  These samples can be used for precision spectroscopy
amplitude analysis. Coupled with the currently available results from
CLEO-c, BES-III will make it possible to study in detail, and with
unprecedented high precision, light hadron spectroscopy in the decays
of charmonium states. In addition, about 90~million $D\bar D$ pairs
will be collected at BES-III in a three-year run at the $\psi(3770)$
peak, which will allow many high precision measurements, including CKM
matrix elements related to charm weak decays, decay constants $f_{D+}$
and $f_{D_S}$, Dalitz decays of three-body $D$~meson decays, searches
for CP~violation in the charmed-quark sector, and absolute decay
branching fractions.  With modern techniques and huge data samples,
searches for rare, lepton-number-violating, flavor-violating, and/or
invisible decays of $D$~mesons, charmonium resonances, and $\tau$~leptons
will be possible.

 Since 2010 experiments have been in progress at
the upgraded VEPP-2000 $e^+e^-$ collider operated in the
center-of-mass energy range from the threshold of hadron production up
to 2\,GeV.  Two detectors are used: CMD-3 and SND.  The goal of the
CMD-3 and SND experiments is to study the spectroscopy of the light
vector mesons ($\rho$, $\omega$, and $\phi$ and their excitations) and to
measure the cross sections of various exclusive channels of $e^+e^-$
annihilation with high accuracy. Such measurements should help clarify
the muon $g-2$ puzzle and provide opportunities for detailed studies
of the dynamics of multi-hadron final states. The expected data
samples should be sufficiently large to disentangle various
intermediate mechanisms, such as those present in the first
high-statistics studies of the four-pion final state in $e^+e^-$
annihilation at CMD-2~\cite{Akhmetshin:1998df,Akhmetshin:2004dy} and
$\tau$ decays at CLEO~\cite{Edwards:1999fj}. A crucial issue for a
successful partial-wave analysis is to use full event
information. With future running in the energy range from 1 to 2\,GeV
one expects data samples of $10^5$ and larger for the dominant final
states with three to six pions.

 The VEPP-4M $e^+e^-$ collider covers a center-of-mass energy range from 2\,GeV to 11\,GeV. It is currently operated in
the charmonium family range with the KEDR detector. Successful
application of two methods for the high-precision determination of the
absolute beam energy---resonant depolarization and Compton
backscattering---resulted in various experiments with record
accuracy. Among them are measurements of the $J/\psi$ and $\psi(2S)$
masses~\cite{Aulchenko:2003qq}; of the total and leptonic width of the
$J/\psi$~\cite{Anashin:2009pc}, $\psi(2S)$~\cite{Anashin:2011ku}, and
$\psi(3770)$~\cite{Anashin:2011kq}; the $D^0$ and $D^\pm$
masses~\cite{Anashin:2009aa}; the $\tau$ lepton
mass~\cite{Eidelman:2011zzb}; and a search for narrow resonances from
1.85\,GeV to the $J/\psi$ mass~\cite{Anashin:2011xm}. Also planned is a
new measurement of $R$ up to 8\,GeV.

The Belle-II experiment~(KEK, Japan) will build off the enormous success of the previous BaBar~(SLAC, USA) and Belle experiments.  The primary goal will be to use $e^{+}e^{-}$ collisions to produce correlated pairs of $B$~mesons through $\Upsilon(4S)$ decays.  As demonstrated at BaBar and Belle, these decays allow precision tests of the Standard Model.  But at the same time, decays of $B$~mesons have proven to be an efficient source of many of the still-unexplained $XYZ$ states of charmonium.  In addition, other techniques, such as initial-state radiation, in which the center of mass of the $e^{+}e^{-}$ annihilation is lowered via the radiation of an initial-state photon, have also allowed for the discovery of other $XYZ$ states, as well as the production of light-quark vector mesons.  While BaBar and Belle accumulated of the order of 1.5~ab$^{-1}$ of data, Belle-II will use the upgraded collider at KEK to collect a projected 50~ab$^{-1}$ of $e^{+}e^{-}$ data by 2020.

 The LHCb experiment~\cite{Alves:2008zz} is designed to exploit the
huge $b\overline{b}$ cross section at $pp$ collisions at LHC
energies~\cite{Aaij:2010gn} for precision flavor physics. The same
characteristics that optimize LHCb for $b$ physics also make it an
excellent charm physics experiment, benefiting from a charm cross
section of $\left(6.10\pm 0.93\right)\,\mathrm{mb}$ in
$7\,\mathrm{TeV}$ proton--proton collisions as obtained in Ref.~\cite{LHCb:2010lga} by extrapolating measurements reported in Ref.~\cite{Aaij:2013mga} using PYTHIA. This
leads to enormous, and still growing, data sets of beauty and charm
hadrons, with tens of millions of clean signal events. Such
high-statistics data samples constitute a huge opportunity for high-precision flavor physics, but they also challenge the theoretical
tools we have to analyze these data sets, including Dalitz plot and partial-wave
analyses.

\subsection{Current analysis techniques}
\label{sec:anal-tech}

\subsubsection{Diffraction and diffractive dissociation}  
\label{sec:analysis.diffraction}
\indent

\begin{figure}
  \begin{center}
      \includegraphics[width=0.7\textwidth]{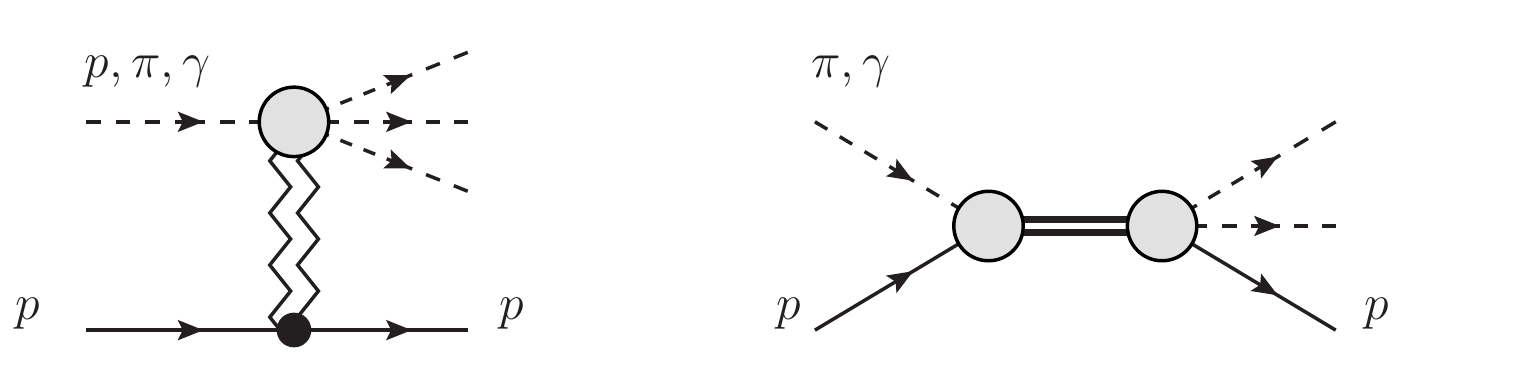}
     \end{center}
\caption{Schematic representation of diffractive dissociation (left) and direct production (right).  \label{fig:BeamFragDirectProd} }
\end{figure}

These processes---see Fig.~\ref{fig:BeamFragDirectProd} (left)---refer to 
production of resonances from dissociation of
either the beam or the target. In a high-energy collisions these can be
kinematically separated due to a large rapidity gap. The following
discussion focuses on meson production from dissociation of a high-momentum meson (pion) beam in the hadronic or Coulomb field of the  
target nucleus.  The methodology equally well applies to the target
fragmentation region.

The analysis typically starts with the selection of events
corresponding to a given final state. Exclusive events are selected by
imposing 
energy conservation between incoming and outgoing particles and 
transverse momentum balance between incoming and outgoing particles
including the recoil particle.   

To disentangle the resonances contributing to a given final state, a
partial-wave analysis (PWA) is performed, which involves certain
model assumptions.   
At high $\sqrt{s}$, the reaction can be assumed to proceed via $t$-channel
exchange between the target and the projectile, which excites the
projectile to a state $X$ and leaves the target intact. The state $X$ then decays into 
a multi-particle final state without further final-state
interaction. This sequence is described by the 
phenomenological approach  of the isobar model~\cite{Salgado:2013dja}. 
In this model the production and the
decay of a state $X$ with quantum numbers $\chi = I^G(J^{PC})M$ 
factorize into a production amplitude $T_{\chi r}(t',s,m)$ and a decay
amplitude $A_{\chi \zeta}(\tau,m)$, where $r$ summarizes the quantum numbers
(e.g.\ helicities) of the projectile, the target, and the recoil, $t'$
and $s$ are the reduced 4-momentum transfer squared and squared center-of-mass
energy, respectively, $m$ is the invariant mass of the intermediate
state $X$, and $\tau$ denotes the set of kinematic variables describing the decay
of $X$ into a particular decay channel denoted by $\zeta$.  
The propagation of the intermediate state $X$ is described by a 
a propagator-like amplitude $K_\chi(t',s,m)$, which 
carries a phase and depends on $m$. 
The amplitude for observation of a given final state is then written 
as
\begin{equation}
  \label{eq:amplitude}
  \mathcal{A}_r(t',s,m) = \sum_{\chi\zeta} T_{\chi r}(t',s,m) K_\chi(t',s,m) 
  A_{\chi\zeta}(\tau,m) \,,   
\end{equation}
where the sum runs over all possible partial waves $T_{\chi r}
A_{\chi\zeta}$, i.e.\ quantum numbers of intermediate 
states $X$ and decay channels $\zeta$ which 
lead to a given final state observed in the experiment. 

Usually, the analysis is performed in a two-step approach. In the
first step, the data are partitioned in small bins of $m$ and
$t'$, such that the production amplitudes can be assumed to be constant
within each bin. 
For a fixed beam energy, i.e.\ $s$ fixed, and $s\gg t'$, the propagator
$K$ can be 
assumed to be approximately 
constant in a small mass bin and will hence be absorbed into the production
amplitude.     
The
amplitude for fixed $m$ and $t'$ is then written as 
\begin{equation}
  \label{eq:amplitude1}
  \mathcal{A}_r = \sum_{\chi\zeta} T_{\chi r}
  A_{\chi\zeta}(\tau)\,.    
\end{equation}

The decay amplitudes $A_{\chi\zeta}(\tau)$ can be calculated 
using the isobar model, in which the decay of $X$ is 
described as a series of   
sequential two-body decays into intermediate resonances (isobars), 
which eventually decay into the final state observed in the
experiment. The sequence of two-body decays is calculated using a
suitable spin formalism~\cite{Chung:1971ri}, taking 
into account isospin and Bose symmetry. 
An important feature of the two-body amplitudes is the factorization
into an angular part, described e.g.\ by Wigner-$D$-functions, and a
dynamical part. The dynamical part contains the respective
Clebsch--Gordan coefficients for a given two-body decay, a dynamical
function parameterizing the dependence of the amplitude on the mass of 
the decaying state (``isobar parameterization''), and the partial-wave
decay amplitude (sometimes 
also called ``isobar factor'') which
represents the overlap of the wave function of the mother 
particle with the two-body wave function of the daughter
particles. In the analysis framework used, these amplitudes are
assumed to be independent of the two-body kinematics, and are thus 
absorbed into the respective production amplitudes, which then depend
not only on the quantum numbers $\chi$ of $X$, but also on the
specific decay channel $\zeta$, summarized as $\xi\equiv
\chi\zeta$ in $T_{\xi r}$.
\footnote{Note that the amplitudes $T_{\xi r}$ and $A_\xi$ are now different from the ones in Eq.~\eqref{eq:amplitude1}, because the isobar factors have been moved from the decay amplitudes to the production amplitudes. To avoid unnecessary clutter, we use the same symbol for both amplitudes.}
 
Complications to the isobar model arising from unitarity constraints 
are difficult to treat theoretically and 
are usually neglected with the argument that normally not all  
possible decay modes of the intermediate state $X$ are
fit simultaneously, but only a small subset or even a single final
state is considered.  
Unitarity in the two-body subsystems is also an issue if
more than one 
(narrow) isobar is included for a given state $X$ because then the
partial-wave decay amplitudes may no longer be considered constant, but will
depend on the two-body kinematics.   
An analysis of 
the reaction $\pi N\rightarrow \pi\pi N$  
incorporating unitarity in the two-body isobar channels using dispersion
relations~\cite{Aitchison:1979fj} has come to 
the conclusion that the results are compatible with an analysis
ignoring the constraints~\cite{Aitchison:1979gq}. With the much larger
data sets available today, 
this issue is certainly worth being revisited. 
Current work which focuses on implementing unitarity constraints in the isobar model are summarized in Sec.~\ref{sec:theory}.

The isobars are typically parameterized using relativistic
Breit--Wigner functions with mass-dependent widths (if branching ratios
are known) or a Flatt\'{e} ansatz. It is well known that this ansatz
is not justified for the $\pi\pi$ scalar-isoscalar partial wave. Different authors
provide different parameterizations for this important amplitude, see, e.g., Ref.~\cite{Au:1986vs,Anisovich:2002ij}. A lot of progress
 has been achieved  towards a full theoretical control of  this channel in recent years, cf.\  Sec.~\ref{sec:theory}. 

A new method
to determine the isobar dynamics directly from the data, 
developed in the framework of the COMPASS experiment, gives very
promising results, but is at the moment limited to few isobars only 
due to a drastic increase of fit parameters. 

The observed multi-differential angular distribution  
is written as the coherent sum over all partial-wave amplitudes
leading to the same final state, 
\begin{equation}
  \label{eq:intensity}
  \frac{\mathrm{d}\sigma}{\mathrm{d}\tau} \propto
  \sum_r \bigg| \sum_{\xi} T_{\xi r} A_{\xi}(\tau) \bigg|^2 \equiv I\,.
\end{equation}
This general form also includes a sum over $r$,
which allows for 
possible sources of incoherence in 
the production process, e.g.\ due to unobserved helicities of incoming
particle, target or recoil, but also due to experimental
effects such as finite resolutions. 
Defining the elements of the spin-density matrix $\rho$ as 
\begin{equation}
  \label{eq:spin-density}
  \rho_{\xi\xi'}=\sum_r T_{\xi r}T^\ast_{\xi' r}\,,
\end{equation}
the angular distribution Eq.~\eqref{eq:intensity} can be also written as
\begin{equation}
  \label{eq:intensity2}
  I = \sum_{\xi\xi'} \rho_{\xi\xi'} A_\xi(\tau) A^\ast_{\xi'}(\tau)\,.
\end{equation}
Based on the observed angular correlations of the final-state
particles, the intensity in each bin is thus decomposed into partial waves
with definite spin and parity, without any prior
assumptions on the shape of the amplitude as a function of $m$ or
$t'$.\footnote{If the data are not sufficient to make fine bins in
  both $m$ and 
  $t'$, the dependence of a given partial wave on $t'$ is
  taken into account by specific functions $f_\xi(t')$.}  

The analysis is commonly performed in a reference frame, in which the
$y$ axis is normal to the production plane, and the $z$ axis, i.e.\
the quantization axis, is chosen along some preferred direction in the
production plane, e.g.\ the direction of the beam or the recoil
particle. In such a system, 
parity conservation in the scattering process is conveniently taken
into account by using a basis of states with a definite symmetry under a
reflection through the production plane, given by the reflectivity
$\epsilon$, and defined as
\cite{Chung:1974fq} 
\begin{equation}
  \label{eq:reflectivity}
  \ket{JM;\epsilon} = c(M) \left[ \ket{JM} -\epsilon P (-1)^{J-M}\ket{J\,-\!M}
  \right]\,,
\end{equation} 
with $\epsilon=\pm 1$ for bosons (mesons), 
$c(M>0)=1/\sqrt{2}$, $c(M=0)=1/2$, and $M$ the modulus of the spin
projection onto a given axis. 
The reflectivity is  defined such that, for beams with pseudoscalar particles, it 
corresponds to the naturality of
the Regge trajectory exchanged in the scattering process.
Parity conservation implies that states with different reflectivity do
not interfere, and the intensity can then be written as 
\begin{equation}
  \label{eq:intensity3}
  I = \sum_\epsilon \sum_{\xi\xi'} \rho_{\xi\xi'}^\epsilon
  A_\xi^\epsilon(\tau) A^{\epsilon\ast}_{\xi'}(\tau)\,. 
\end{equation}

In principle an infinite number of waves is needed in the partial-wave
expansion of the
cross section~\eqref{eq:intensity3}. A limited amount of data,
however, requires a 
truncation of the series and hence a possibly biased selection of
waves that are included in the fit. Larger data sets 
help to reduce the model bias because more waves can be included in
the fit. A model-independent algorithm for the selection of the wave
set, based on genetic evolution, was introduced in~\cite{Neubert:2012}. Also this algorithm, however, requires tuning
of parameters and may hence contain some residual bias, but it is
definitely superior to the standard approach of selecting waves ``by
hand.'' 
The fit normally also contains a background wave, characterized by a 
uniform distribution in $n$-body phase space, which is added
incoherently to the other waves.  The need for the 
inclusion of those background terms is a clear indication that the
amplitudes used are incomplete and further theoretical progress
is necessary---some possible routes for future developments
are outlined in Sec.~\ref{sec:theory}.

In the first step of the analysis, an event-based
extended log-likelihood 
fit of the probability density in the full final-state
phase space is usually performed to determine the 
complex production amplitudes $T_{\xi r}$ 
for each bin 
of the final-state invariant mass $m$ and of the 4-momentum
transfer $t'$.  
The elements of the spin-density matrix are then calculated as
$\rho_{\xi\xi'}^\epsilon=\sum\nolimits_r T_{\xi r}^\epsilon T_{\xi'
  r}^{\epsilon\ast}$. 
The diagonal elements of the spin-density matrix are the intensities
of the corresponding waves, while the off-diagonal elements determine
the phase differences between two waves. 
The fit also takes into account the experimental
acceptance of the spectrometer, calculated from a phase-space Monte
Carlo simulation of the 
apparatus. 

The result of the first step of the PWA is an
independent spin-density matrix for each
$m$ and $t'$ bin, containing all waves used in this particular
bin. 
Apart from reasons of model complexity and computing resources, the
splitting in two steps has the advantage that no dependence of the 
amplitude on the mass $m$ is introduced in the first step. 
Therefore, apart from the assumed factorization into production and
decay amplitudes, no 
assumptions about the resonances in the analyzed $n$-body system enter
the analysis at this point, so that model dependence is kept at a
minimum.  

In the second step, a model is applied in a
$\chi^2$ fit to  
describe the mass and $t'$-dependence of these
matrices, where for computational reasons only a few waves are
considered. For each wave the model includes  
resonant contributions, usually parameterized in terms of relativistic
Breit--Wigner functions with dynamic widths and parameters
{\it independent} of $t'$, 
and non-resonant 
contributions added coherently, in most cases parameterized by
empirical functions. 

\subsubsection{Direct-channel production}
\indent

This refers to a situation when the center-of-mass energy of colliding particles
coincides with a mass of nearby resonances, see Fig.~\ref{fig:BeamFragDirectProd} (right).  
Resonance decay is
analyzed following the same ideas of isobar factorization discussed
above.  Direct-channel baryon excitations decaying to at most two
pseudoscalar mesons and a baryon have often been analyzed by fitting
model amplitudes to the partially integrated differential cross
section and polarization observables. With new, high-statistics data
now becoming available, an event-by-event analysis, similar to the one
described above, ought to be performed.

Most analyses have so far focused on decay channels containing a single
meson and a baryon. Model amplitudes are often based on effective
Lagrangians which incorporate low-energy, e.g.\ chiral and
unitarity constraints.  Two-meson production, especially $\pi\pi$
production, has the largest cross section for photoproduction from
energies around the second resonance region and higher and, since it
couples strongly to many baryon resonances, it is an important channel
that complements the information obtained in single pseudoscalar meson
photoproduction.

There are two main approaches to analyze data on pion or photo-induced 
production of single mesons.  The first approach is the so-called
energy-independent approach or single-energy solution. Here, data on
differential cross sections and  polarization observables are analyzed in fixed
bins of energy (and ideally in fixed bins of solid angle) to extract the
scattering or production amplitudes. When fixed bins in energy {\it and} solid
angle are chosen, the so-called CGLN amplitudes emerge.  For the definition of the CGLN amplitudes and their partial wave decompositions for pion--nucleon scattering~\cite{Chew:1957zz} and photoproduction of a pseudoscalar~\cite{Chew:1957tf}, we refer the interested readers to the original publications. The model-independent determination of these amplitudes is the problem of the so-called {\it complete
experiment} that has attracted much attention lately~\cite{Sandorfi:2010uv,
Wunderlich:2013iga, Workman:2011hi}, including a dedicated
workshop~\cite{Trento2014}. See Ref.~\cite{Tiator:2012ah} for a summary.

The angular  dependence of the CGNL amplitudes yields in principle the
photoproduction multipoles. The main problem is an undetermined overall phase of
the multipoles, that is different not only for different energies but also for
different angles. To disentangle multipoles, usually a truncation in multipoles
is performed, or high partial waves from an existing analysis are held fixed, or
a penalty function is used to guide the single-energy multipole extraction by an
energy-dependent solution. Practically, the precision of existing data poses the
major problem in the extraction of amplitudes. Single-energy solutions have been
extracted for many years~\cite{Arndt:2006bf,Workman:2012jf,Drechsel:2007if}.

For pseudoscalar meson scattering, theoretically, only three observables need
to be measured with good precision to reconstruct the scattering amplitudes up
to one common phase. For the reaction $\pi^- p\to K^0\Lambda$, spin rotation
parameters have been  determined even though with limited solid-angle coverage.
A Bonn--Gatchina fit to the data provided a set of single-energy solutions (6 per
energy point)~\cite{Anisovich:2013tij}. The solution closest to the 
energy-dependent solution was chosen as the physical one. This solution  was
smooth and very close to the energy-dependent solution. The second best 
solution showed larger fluctuations in all partial waves. 

The situation is much more complicated in the case of photoproduction
amplitudes. Here at least eight observables are needed for an unambiguous
extraction of the amplitudes. The single-energy approach  is particularly suited
for the analysis of reactions with  kaon--hyperon final states, where the recoil
polarization of the final baryon can be measured. Another possibility is the
analysis of data on meson production taken at relatively low energies where the
number of contributing multipoles is restricted and only few observables are
needed to construct the amplitudes. In~\cite{Hartmann:2014mya}, this  method was
exploited to determine the helicity amplitudes of the $N(1520) 3/2^-$ resonance
from an energy-independent analysis of $S$-, $P$-, and $D$-waves in $\gamma p\to p\pi^0$. 

The second approach is the so-called energy-dependent partial-wave analysis.
Here the angular and energy dependencies are analyzed simultaneously. A weak
point of this approach is the large number of parameters and the large amount of
computer time to obtain a solution. However, this approach offers also several
advantages. First of all, global properties of the amplitudes---like unitarity
and analyticity---can be imposed right from the beginning. The  unitarity
property is a powerful constraint when elastic and inelastic  channels are
analyzed jointly. As mentioned, a given resonance has one pole position, which
is fixed in all reactions. Its couplings to  the different channels define its
contributions to the different pion-  and photo-induced reactions.
Energy-dependent coupled-channel analyses are thus the method of choice to
search for weak resonance signals. Single-energy solutions, on the other hand,
can be used to search for very narrow structures that tend to be missed by
energy-dependent analyses~\cite{Arndt:2006bf}.

At this point it should be stressed that better data for pion-induced reactions
are of urgent need. In coupled-channel analyses, these data determine the
hadronic part of the amplitude and consequently also the photo- and
electroproduction amplitudes. However, many measurements of pion-induced
reactions date back to twenty or more years ago. Often, systematic uncertainties
are not reported, or are known to be underestimated. The need for hadronic beams
has been recently discussed on a dedicated workshop~\cite{hadbeam}.

Groups such as SAID, MAID, J\"ulich/Athens/GWU, EBAC (now Argonne--Osaka), Gie{\ss}en,
and Bonn--Gatchina have all made valuable contributions to this field,
and anticipate being able to utilize further data as it becomes
available.

In the following Chapter we detail some of the theoretical constraints
that are being developed and that need to be implemented to verify the
various assumptions of these analysis techniques.

\section{Amplitude Analysis}
\label{sec:theory}
\vskip-5pt [C.~Hanhart] \vskip10pt

Hadron spectroscopy aims at the identification of hadron resonances
and the determination of their properties. In the limit of a large
number of colors, hadrons become bound states of constituent
quarks. In reality, almost all of them are resonances that decay
strongly to ground state hadrons---pions, kaons, etas, and
nucleons. The heavier the resonance, the more multi-particle channels
are allowed kinematically as final states. As a result, resonances
become broad, overlap, and their identification gets increasingly
difficult. The goal of the amplitude analyses outlined here is to pin
down the spectrum in the so called resonance region which typically
corresponds to excitation energies not greater than $2$--$3$\,GeV.

The easiest and most commonly used parametrization for decays and
scattering amplitudes is built from sums of Breit--Wigner functions
(BWs) with energy-dependent widths, sometimes accompanied by
(smoother) background terms. While this ansatz typically allows for a
high-quality fit of many-body final states, it suffers from various
problems. The poles of the BWs are in general not identical to the
true poles of the $S$-matrix. As such their parameters may differ
between different reactions, which prevents a systematic, consistent
study of many final states.  Typically BWs do not  reproduce the analytical properties of  reaction amplitudes.  In addition sums of BWs  violate unitarity. For instance, in the case 
of $2\to 2$ scattering,  unitarity  correlates the energy-dependent complex phases
between the different BWs.  In decays with only two strongly-interacting particles in the elastic regime,  Watson's theorem imposes the equality between scattering and production
phases.  Therefore sums of BWs may only be a valid approximation when considered far from kinematic thresholds and only for poles close to the real axis that  are far from each other, that is for narrow and isolated resonances.

In this section we outline theoretical aspects that need to be
considered to arrive at parameterizations of amplitudes that try to
minimize the effect of the above-mentioned problems.
From the point of view of reaction theory, also known as $S$-matrix
theory, resonances are poles of partial-wave scattering amplitudes in
the unphysical domain of kinematical variables, energy, and/or angular
momenta. Thus, their identification requires an analytic extension of the (multi-channel) amplitudes into the complex plane of the kinematic variables.  $S$-matrix theory imposes severe constraints on the amplitudes allowed, such as unitarity, analyticity, as well as
crossing symmetry.  In addition, the amplitudes have to be consistent
with the assumed discrete symmetries of the underlying theory.
Depending on the kinematical regime of an experiment different aspects
of this list may become relevant. For example, low-energy scattering
is dominated by a few elastic partial waves, which may be constrained
by unitarity, analyticity, and in some special cases crossing symmetry
(cf.\ Sec.~\ref{sec:disp} on dispersion theory).  To control
subleading singularities, or if there is no sufficient information
about particle scattering available to employ dispersion theory, in
addition to the general principles, it is sometimes necessary to
impose further properties on the reaction dynamics, e.g.\ from
long-ranged meson exchanges whose strength may be constrained from
data (e.g.\ the strength of the pion exchange in $\pi\rho\to \rho\pi$
is given by the width of the $\rho$~meson) or by chiral symmetry (cf.\
Sec.~\ref{sec:coup_chan} on dynamical coupled-channel methods and
related approaches). On the other hand, a detailed understanding of
resonance production with high-energy beams may require knowledge of
singularities in the complex angular momentum plane---Reggeons (cf.\
Sec.~\ref{sec:dual} on duality and finite-energy sum rules).

In general, amplitude analysis can be considered as a three-step
process. In step one, theoretical amplitudes are proposed and
constrained by fitting the experimental data. In step two, these
amplitudes are tested against various constraints that are used to
minimize the amount of unresolved ambiguities in the amplitude
determination.  Finally in step three, the amplitudes are extrapolated
(analytically continued) to the unphysical kinematical region of
energy and angular momentum to determine properties of resonances.

With the advent of new high-statistics experiments, combined with the
development of theoretical tools, the widely used isobar model can
now be replaced by model-independent analyses. Connecting the emerging
lattice results with the parameters extracted using the analysis
techniques mentioned above will provide a direct contact between
experimental data and QCD.

\subsection{Dispersive methods} 
\vskip-5pt [C.~Hanhart, M.~Hoferichter, B.~Kubis, J.~R.~Pel{\'a}ez] \vskip10pt
\label{sec:disp}

In this section we will discuss several examples where dispersion
relations (DRs) have been applied with the aim of obtaining precision
parameterizations of amplitudes at low energies and performing their
analytic continuation. Another important aspect that concerns the
connection of low-energy physics and the high-energy region within
dispersion theory will be touched upon in Sec.~\ref{sec:dual}.

A resonance is uniquely characterized by its pole and residues, the
position of the pole being universal, its residues depending on the
decay channel in question.  The challenge in the precision
determination of these parameters lies in the restriction that
experiments are limited to real, physical values of the center-of-mass
energy $s$. In principle, DRs provide a rigorous way of analytically
continuing amplitudes from the physical regime into the complex plane,
and thus of unambiguously extracting the pole parameters of the
resonance.  Only when a resonance is well isolated from others and is
also far from thresholds, one can use simple expressions like
Breit--Wigner amplitudes that provide, in a limited region, a very
good approximation to the result one would obtain from dispersion
theory. Mathematically, these are cases where the distance of the
resonance pole to the real axis is smaller than its distance to any
other singularity, or where there is just one threshold cut
nearby. Resonances corresponding to such a situation have been
thoroughly studied and their properties are well established.
Nowadays we are trying to understand the complicated part of the
spectrum, where this ideal situation often does not occur and
resonances are wide, with poles relatively deep in the complex
plane. Effects of overlapping resonances and proximity to more than
one threshold due to many possible decay channels require more
elaborate techniques.

For a general introduction to dispersive techniques, we refer to
Refs.~\cite{HaMa,HaMa2,HaMa3}. Briefly, in terms of physics, DRs are a
consequence of causality, which mathematically allows us to
analytically extend the amplitudes into the complex plane, and then
use Cauchy's theorem to relate the amplitude at any value of the
complex plane to an integral over the (imaginary part of the)
amplitude evaluated on the real axis, where data are available. Such a
relation can be used in several ways. On the physical real axis, it
implies that the amplitude has to satisfy certain integral
constraints. Thus, one can check the consistency, within
uncertainties, of the data at a given energy against the data that
exist in other regions. Additionally, DRs may be imposed as
constraints, by forcing the amplitude to satisfy the DR while fitting
the data.  Finally, certain sets of coupled DRs are so strongly
constrained (see the discussion of Roy equations below) that they can
actually be \textit{solved} as a boundary problem in a limited
(typically low-)energy range, given a specific high-energy input and
depending on a well-defined number of parameters (subtraction
constants)~\cite{Gasser:1999hz,Wanders:2000mn,Ananthanarayan:2000ht,Buettiker:2003pp}.

Especially, one can even use a DR to obtain values for the amplitude
at energies where data do not exist, using existing data in other
regions. Once one has an amplitude that satisfies the DR and describes
the data well, it is possible to extend the integral representation to
obtain a unique analytic continuation into the complex plane (or at
least to a particular region of the complex plane where the validity
of the DR can be rigorously established).  For partial-wave
amplitudes, one can thus study the complex-energy plane and look for
poles and their residues, which provide the rigorous and
observable-independent definition for the resonance mass, width, and
couplings.

Prime examples for precision determinations of resonance pole
positions by dispersive techniques concern the $\sigma$ or
$f_0(500)$~\cite{Caprini:2005zr,GarciaMartin:2011jx} as well as the
$\kappa$ or $K_0^*(800)$ resonance~\cite{DescotesGenon:2006uk}.  While
both are still ``simple'' in the sense that they are overwhelmingly
dominantly \textit{elastic} resonances (in $\pi\pi$ and $\pi K$
scattering, respectively), their poles are non-trivial to determine
since they lie far away from the real axis, with widths of about
550\,MeV in both cases. By convention, the width $\Gamma$ of a
resonance is defined as $\Gamma=-2\,\text{Im}\sqrt{s_p}$, where $s_p$
denotes the complex pole position of the resonance.  The (complex)
range of validity of the corresponding DRs is restricted by the
singularities of the so-called double-spectral region, as well as by
the requirement of the partial-wave projection to converge, and can be
shown to still comprise the poles under investigation.  One
furthermore employs the consequence of unitarity that poles of the $S$-matrix on the
second Riemann sheet correspond to zeros on the first sheet; the
positions of the latter are determined in practice.  As the partial
waves in these cases are given by DRs using imaginary parts along the
real axis only, with kernel functions known analytically, this
procedure is then straightforward.

DRs have been extensively studied for various $2\to 2$ reactions, with
a few extensions to include more complicated final
states~\cite{Hoyer}.  Amplitudes for two-body reactions depend on the
Mandelstam variables $s$ and $t$ (or $u$), which are related to
center-of-mass energy and momentum transfer, respectively. Typically,
DRs are formulated in terms of $s$, with the $t$-dependence either
fixed or integrated over.  The former are referred to as ``fixed-$t$
DRs.'' Of special importance among these kinds of DRs is the case
$t=0$ for elastic reactions, known as ``forward DRs,'' since, due to
the optical theorem, the imaginary part of the forward amplitude is
proportional to the total cross section, and data on total cross
sections are generically more abundant and of better quality than on
amplitudes for arbitrary values of $s$ and $t$.

On the other hand, one can eliminate $t$ by projecting the amplitude
onto partial waves, for which then a DR is written. The advantage of
these partial-wave DRs is that their poles on the second Riemann sheet
are easily identified as resonant states with the quantum numbers of
the partial wave. Therefore, they are very interesting for
spectroscopy. However, due to crossing symmetry, partial waves have a
left-hand cut in the unphysical $s$ region, which also contributes to
the DR. If the region of interest lies very far from this cut, it can
be neglected or approximated, but when closer, or if one wants to
reach a good level of precision, it becomes numerically relevant and
has to be taken into account.  Since the amplitude in the unphysical
region may correspond to different processes arising from crossed
channels in other kinematic regions and other partial waves, this
complicates the construction of DRs substantially. Dealing rigorously
with the left-hand cut usually involves an infinite set of coupled
integral equations, known for $\pi\pi$ scattering as Roy
equations~\cite{Roy:1971tc}, but other versions exist for $\pi
K\rightarrow\pi K$, $\gamma\gamma\rightarrow\pi\pi$, and $\pi
N\rightarrow\pi N$, under the generic name of Roy--Steiner
equations~\cite{Steiner:1971ms,Hite:1973pm}.  There is a considerable
and relatively recent progress, as well as growing interest in
obtaining rigorous dispersive descriptions of these
processes~\cite{Ananthanarayan:2000ht,Buettiker:2003pp,Colangelo:2001df,Kaminski:2002pe,Pelaez:2004vs,GarciaMartin:2011cn,Oller:2007sh,Hoferichter:2011wk,GarciaMartin:2010cw,Moussallam:2011zg,Ditsche:2012fv},
which play an essential role when describing final states of almost
all other hadronic strongly-interacting reactions.

In all these variants of DRs the integrals formally extend to
infinity. In order to achieve convergence and also to suppress the
high-energy contribution, one introduces so-called subtractions. In a
subtracted version of a given DR, the integrand is weighted by
additional factors of $1/(s-s_0)$, where $s_0$ is referred to as the
subtraction point, at the expense of introducing a priori undetermined
parameters (subtraction constants). For a $2\to 2$ scattering process
in general two subtractions are required to ensure
convergence~\cite{Froissart:1961ux,Martin:1962rt}, but once- or even
less subtracted relations exist for certain amplitudes.  Subtraction
constants can be constrained by matching to effective field theories,
lattice calculations, or simply fits to data.  For the high-energy
region one typically makes use of Regge theory, which is known to
describe data on, for instance, total cross sections up to very large
energies well. Even if data are not very precise or non-existent,
Regge theory allows for predictions for different processes by
combining the results for well-established reactions by means of 
factorization.  Regge predictions are less robust for the
$t$-dependence of the amplitudes, although if only small $t$ are
required, they provide a reasonable approximation. Simple and updated
Regge parameterizations can be found in the Review of Particle
Physics~\cite{Agashe:2014kda}, except for meson--meson scattering for
which we refer
to~\cite{GarciaMartin:2011cn,Pelaez:2003ky,Halzen:2011xc,Caprini:2011ky}.

Since most hadronic observables involve pions, kaons, or light nuclei
in the final state, at some stage their theoretical description
requires input from elastic $\pi\pi$, $\pi K$, and $\pi N$ scattering
via the so-called Fermi--Watson
theorem~\cite{Watson:1954uc,Fermi:2008zz}.  For processes with only
two strongly-interacting final-state particles, it fixes the phase of
the whole amplitude to that of the hadron pair.  A rigorous dispersive
implementation of this theorem can be achieved via the
Muskhelishvili--Omn\`es (MO) method~\cite{Omnes:1958hv,Muskhelisvili},
where the amplitude is expressed in terms of an Omn\`es factor
uniquely determined by the phase of the scattering process of the
final state. This method is particularly well-suited for the study of
meson form factors, not only of pions, kaons, but charmed $D$~mesons as
well, see for
instance~\cite{Gorchtein:2011vf,Hanhart:2012wi,Ananthanarayan:2011uc,
  Abbas:2009dz, Szczepaniak:2010re,Abbas:2010ns,Guo:2010gx,Hoferichter:2014vra} and
references therein. In addition to the right-hand cut accounted for by
the MO method, the description of production amplitudes involves a
left-hand cut. It should be stressed that the structure of this
left-hand cut is different from the left-hand cut of the pertinent
scattering reaction.

Building upon MO techniques, one may obtain a consistent treatment of
$\pi\pi$ rescattering for more complicated reactions as well, e.g.\
using Khuri--Treiman techniques for three-particle
decays~\cite{Khuri:1960zz}.  If for a given decay the contribution
from the left-hand cut is known to be suppressed, e.g.\ for
$\eta,\,\eta'\to\pi^+\pi^-\gamma$~\cite{Stollenwerk:2011zz}, and can
be expanded in a polynomial, this setup reduces to the original MO
solution, while otherwise coupled integral equations need to be
solved. These integral equations happen to be linear in the
subtraction constants, so that the full solution can be reconstructed
by a linear combination of basis functions that correspond to the
choice of one subtraction constant set equal to $1$ and the others put
to zero. In this way, one obtains a description of the amplitude in
terms of a few parameters which can be determined by comparison to
experiment, see~\cite{Hoferichter:2012pm} for the example of
$\gamma\pi\to\pi\pi$.  For a real decay process, the solution of the
integral equations is further complicated by the analytic properties
of the amplitude, which require a careful choice of the integration
contour in the complex plane. For an application of these methods to
$\eta,\omega, \phi \rightarrow 3\pi$ decays
see~\cite{Kambor:1995yc,Colangelo:2009db,Schneider:2010hs,Niecknig:2012sj,Danilkin:2014cra}.

Watson's final-state theorem as well as the more general consequences
thereof encoded in the use of MO and Khuri--Treiman techniques only
apply in the region of \textit{elastic} unitarity (or at least as long
as inelastic effects are sufficiently small to be negligible).  In
principle, the MO method can be generalized to multiple coupled
channels, provided the corresponding multi-channel $T$-matrix is
known.  In practice, this has been implemented mainly for the case of
the $\pi\pi$ isospin $I=0$ $S$-wave, where the inelasticity sets in
sharply at the $\bar KK$ threshold, which at the same time almost
coincides with the position of the $f_0(980)$ resonance.  In this
case, the additional input needed beyond the $\pi\pi$ scattering phase
shift are modulus and phase of the $\pi\pi\to\bar KK$ transition.
Applications have mainly concerned scalar form factors of different
kinds~\cite{Donoghue:1990xh,Moussallam:1999aq,DescotesGenon:2000ct,Hoferichter:2012wf,Daub:2012mu}.
For the $\pi K$ system, strangeness-changing scalar form factors have
been studied, taking the coupling to $\eta K$ and $\eta' K$ into
account~\cite{Jamin:2001zq,Doring:2013wka}.  It needs to be said,
though, that this method can be realistically applied mainly in
contexts where inelasticities are dominated by one or two channels;
compare also the suggestion to approximate the coupling to additional
channels via resonances only~\cite{Hanhart:2012wi}.  The combination
of the Khuri--Treiman method to treat three-body decays with inelastic
channel coupling has not been undertaken to date.

For more complicated processes a rigorous formulation of DRs soon
becomes extremely demanding.  In such a situation, one could try to
use models that incorporate at least the most relevant analytic
structure, impose further constraints in the form of sum rules, and
make sure that the resonances claimed lie within the applicability of
the approach. Some models, based on simplified DRs, as for instance
the $N/D$ method or some unitarized models, can be very useful to
obtain resonance poles and parameters in cases with coupled channels,
at least in those channels where reliable data exist.  By all means,
one should refrain from making spectroscopic claims from simple models
that fail to obey these constraints.

\subsection{Dynamical coupled channels,
  Chew--Mandelstam, \textit{K}-matrix, and related approaches
  in baryon analyses}
\label{sec:coup_chan}
\vskip-5pt [M.~D\"oring] \vskip10pt

The phenomenology in the analysis of excited baryons is complex. One complication
arises from known strong inelasticities into multi-pion states, mostly $\pi\pi
N$. For example, two pions with the $\rho(770)$ quantum numbers are known to be
responsible for inelasticities at higher energies. The two pions and the nucleon
can also be in relative $S$-wave, i.e.\ one can have the effective quantum
numbers of a $\sigma N$ state. With the centrifugal barrier absent, this
configuration leads to large inelasticities into the $\pi\pi N$ channel even at
low energies, resulting in the unusual resonance shape of the (very light) Roper
resonance $N(1440)1/2^+$. In other words, from the standpoint of meson
spectroscopy, one has maximal contamination from excited baryons, while from the
standpoint of baryon analysis, two-body channels as $\pi N$ need to be
supplemented with three-body states.

Furthermore, a two-particle subsystem of the $\pi\pi N$ system can also contain
resonance singularities. As mentioned, the $\pi\pi$ subsystem can have the
quantum numbers of a $\rho(770)$, coupling to the nucleon with a certain
isospin, total spin, and total angular momentum. In general, more than one
configuration is possible. Those singularities lead to branch points in the
complex plane of the overall center-of-mass scattering energy. These
non-analyticities are located on the same sheets as resonances and can lead to
false resonance signals if not properly taken into account~\cite{Ceci:2011ae}. 
Last but not least, the inelasticities from channels formed by a stable baryon
and a stable meson are important. The prime example is the strong coupling of
the $\eta N$ channel to the $S_{11}$ partial wave, in particular the
$N(1535)1/2^-$.

The complex phenomenology of the baryon resonance region has, so far, hindered
the implementation of the rigorous methods discussed in previous sections. Also,
the search for new baryon resonances usually implies a multi-channel fit to data
of different reactions, to look for resonances that couple only weakly to the
$\pi N$ channel. Recently, experimental activity has focused on photo- and
electroproduction reactions, with a variety of final states such as $\pi N$,
$\pi\pi N$, $\eta N$, $\pi\eta N$, $K\Lambda$, $K\Sigma$, and $\omega N$. As
resonance pole positions are independent of the reaction studied, the
simultaneous analysis of different final states facilitates the search for weak
resonance signals.

\begin{figure}
  \begin{center}
  \qquad\quad
      \includegraphics[trim = 200 200 -20
      0,width=0.46\textwidth]{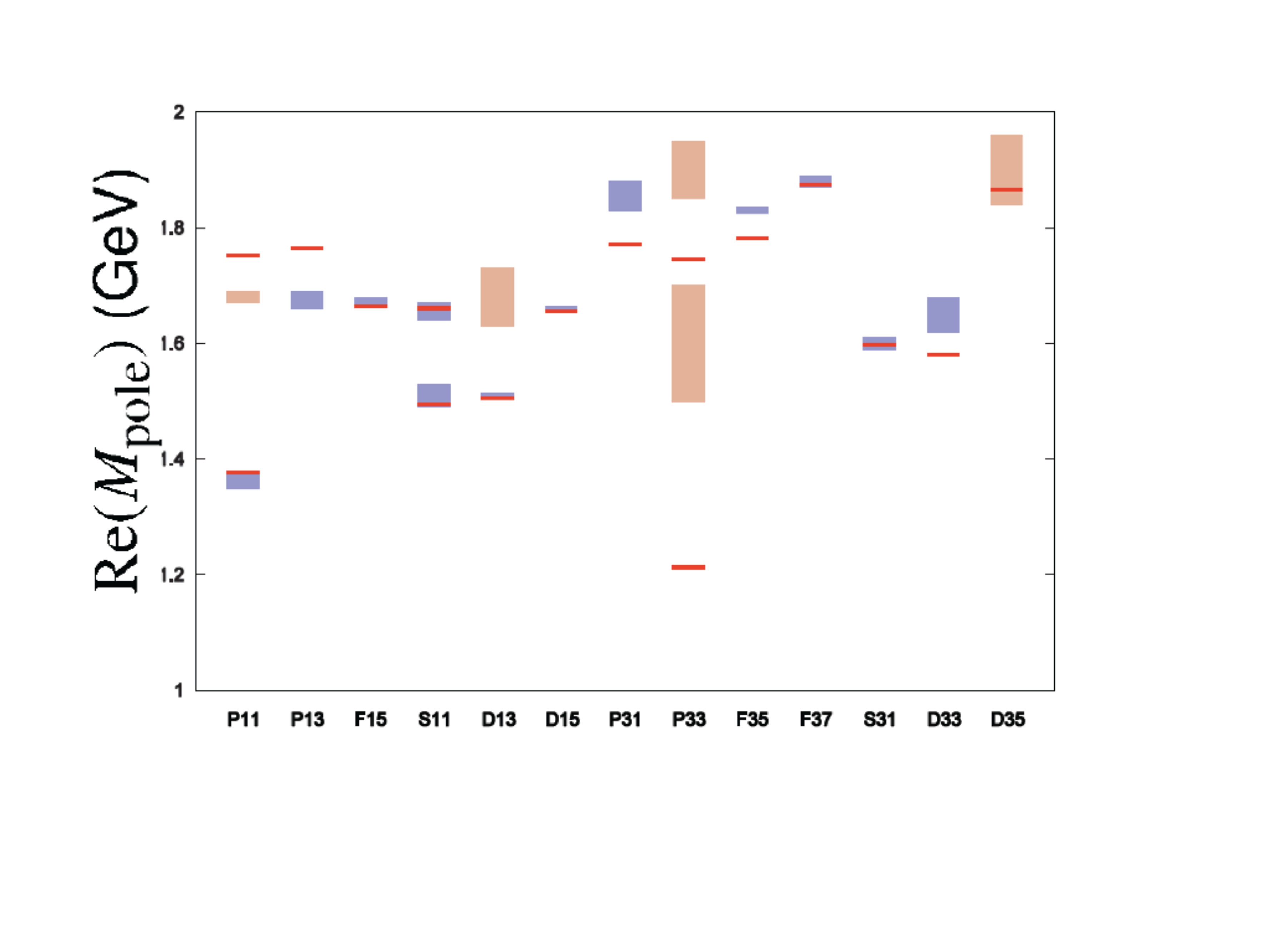} 
   \includegraphics[trim = 200 200 0
      0,width=0.46\textwidth]{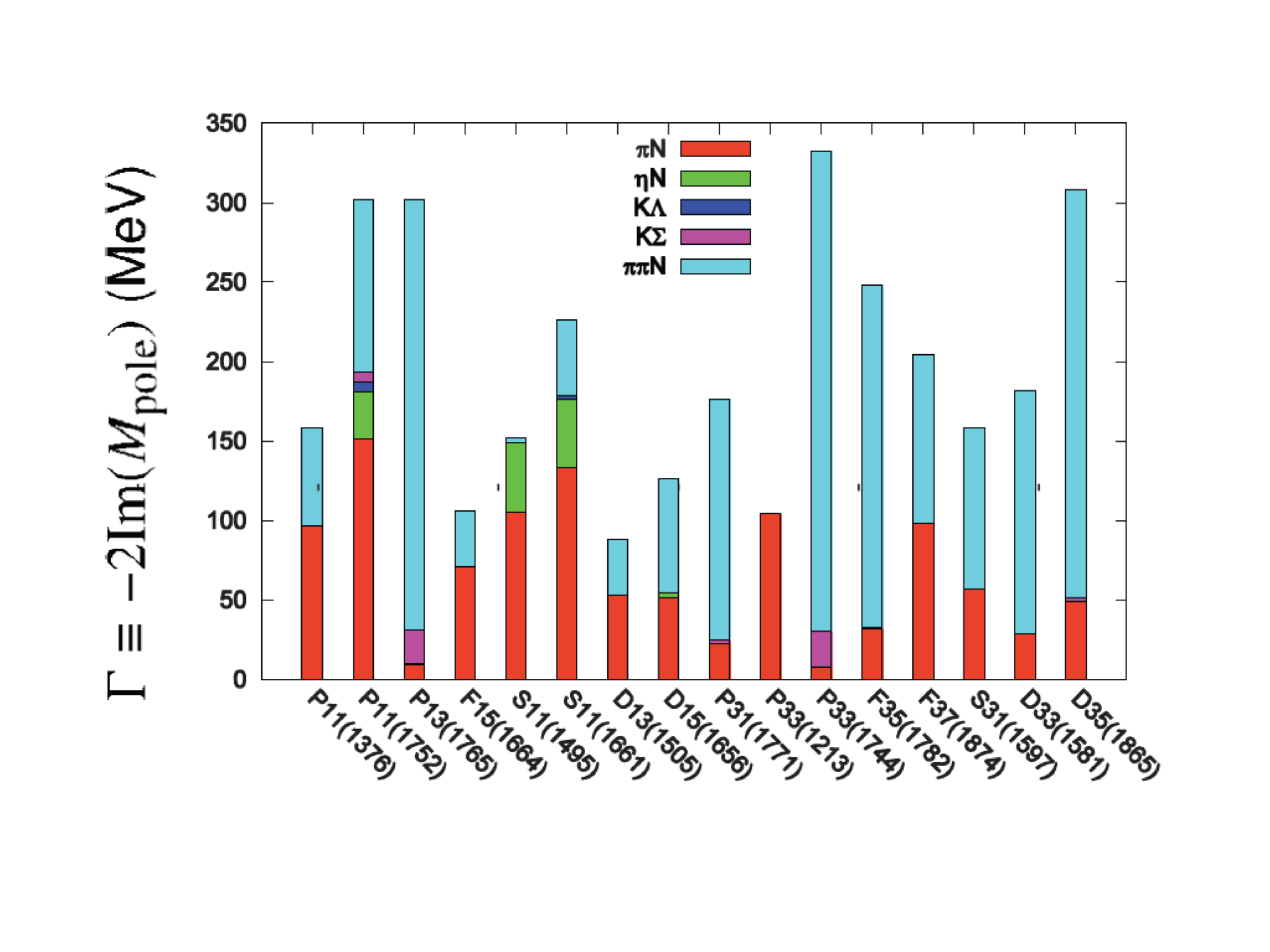}
  \end{center}
\caption{Baryon spectrum: masses (left) and widths (right) from the ANL/Osaka 
approach~\cite{Kamano:2013iva}. \label{fig:ebac}}
\end{figure}
Several analysis tools have been developed for the analysis of excited baryons,
among them the so-called dynamical coupled-channel approaches, pursued in the
ANL--Osaka (former EBAC) collaboration, in the J\"ulich--Athens--GWU collaboration,
in the Dubna--Mainz--Taipeh (DMT) group, and
others~\cite{Kamano:2013iva,Ronchen:2014cna,Ronchen:2012eg,Tiator:2010rp}; 
see Fig.~\ref{fig:ebac} for recent results. The left-hand cuts are approximated
perturbatively by $u$-channel baryon exchanges, while $s$-channel unitarity,
driven by the right-hand cut, is respected exactly. The discussed $\pi\pi N$
three-body states are included such that two-body subsystems describe the
corresponding phase shifts. Subthreshold non-analyticities such as the circular
cut, short nucleon cut, and further left-hand cuts are present. Exchanges in
$t$- and $u$-channel are truncated to the lightest (excited) hadrons. These
exchanges provide a background that connects different partial waves and limits
the room for resonances. In the J\"ulich approach, the $t$-channel dynamics for
the $\rho$ and $\sigma$ quantum numbers is provided by the use of dispersive
techniques and a fit to $N\bar N\to\pi\pi$ data~\cite{Schutz:1994ue}.

Another aspect of three-body dynamics is the consistent implementation of
two-body decays. It has been shown~\cite{Aaron:1969my} that unitarity in the
three-body sense can be achieved by complementing three-body states with
appropriate exchange processes. For example, in the three-pion system a
$\pi\rho(770)[\pi\pi]$ state requires appropriate pion exchanges to fulfill
unitarity. That principle has inspired the construction of dynamical
coupled-channel approaches in baryon analysis as
well~\cite{Kamano:2013iva,Ronchen:2014cna}. In meson analysis, three-body
unitarity has been explored in~\cite{Kamano:2011ih}, using effective Lagrangians
and isobars, that fulfill two-body unitarity and fit the corresponding phase
shifts. If one restricts the rescattering series to the first term, one recovers
an amplitude closely related to the traditional isobar picture that may or may
not be unitary in the two-body sense, but is never unitary in the three-body
sense. Summing up the consistently constructed interaction beyond the leading
term, including rearrangement graphs, restores unitarity in the three-body
sense. See also~\cite{Magalhaes:2011sh}, where three-body unitarity based on
point-like interactions is considered. Coming back to the analysis of excited
baryons, three-body unitarity is sometimes not manifestly included but
effectively approximated by free phases as in the $D$-vector approach by the
Bonn--Gatchina group. Similarly, in the MAID analysis, unitarity is approximated  by
complex phases~\cite{Drechsel:2007if}.

In dynamical coupled-channel approaches, usually a scattering equation with
off-shell dependence of the driving interaction is solved. If the interaction is
factorized on-shell, the integral equation reduces to a matrix equation in
coupled channels. Real, dispersive parts of the intermediate propagating states
can be maintained. Such contributions are relevant for the reliable analytic
continuation to search for resonance poles and residues. 

Analyses of this type are pursued by the GWU/INS (SAID) approach in the
Chew--Mandelstam formulation~\cite{Arndt:2006bf,Workman:2012jf}, by the
Bonn--Gatchina group in the $N/D$
formulation~\cite{Anisovich:2011fc,Anisovich:2012ct}, and by the Kent
State~\cite{Shrestha:2012ep}  and the Zagreb~\cite{Batinic:2010zz} groups in the
Carnegie--Mellon--Berkeley (CMB) formulation.  The Gie{\ss}en group uses a $K$-matrix
formalism~\cite{Shklyar:2004ba,Shklyar:2012js}, while the MAID approach employs a
unitary isobar approach in which the final-state interaction is taken from the
SAID approach~\cite{Drechsel:2007if}. Dispersive approaches and unitary isobar
analyses on meson electroproduction have been performed by the JLab
group~\cite{Aznauryan:2012ba}.  Two-pion electroproduction is analyzed at JLab
as well~\cite{Aznauryan:2005tp}.

In the GWU/INS (SAID) approach the interaction is parameterized without the need
of explicit resonance propagators~\cite{Workman:2012jf}. Resonance poles are
generated only if required by data, which makes this approach particularly
model-independent for baryon spectroscopy.

\begin{figure}[t]
  \begin{center}
    \includegraphics[width=\textwidth]{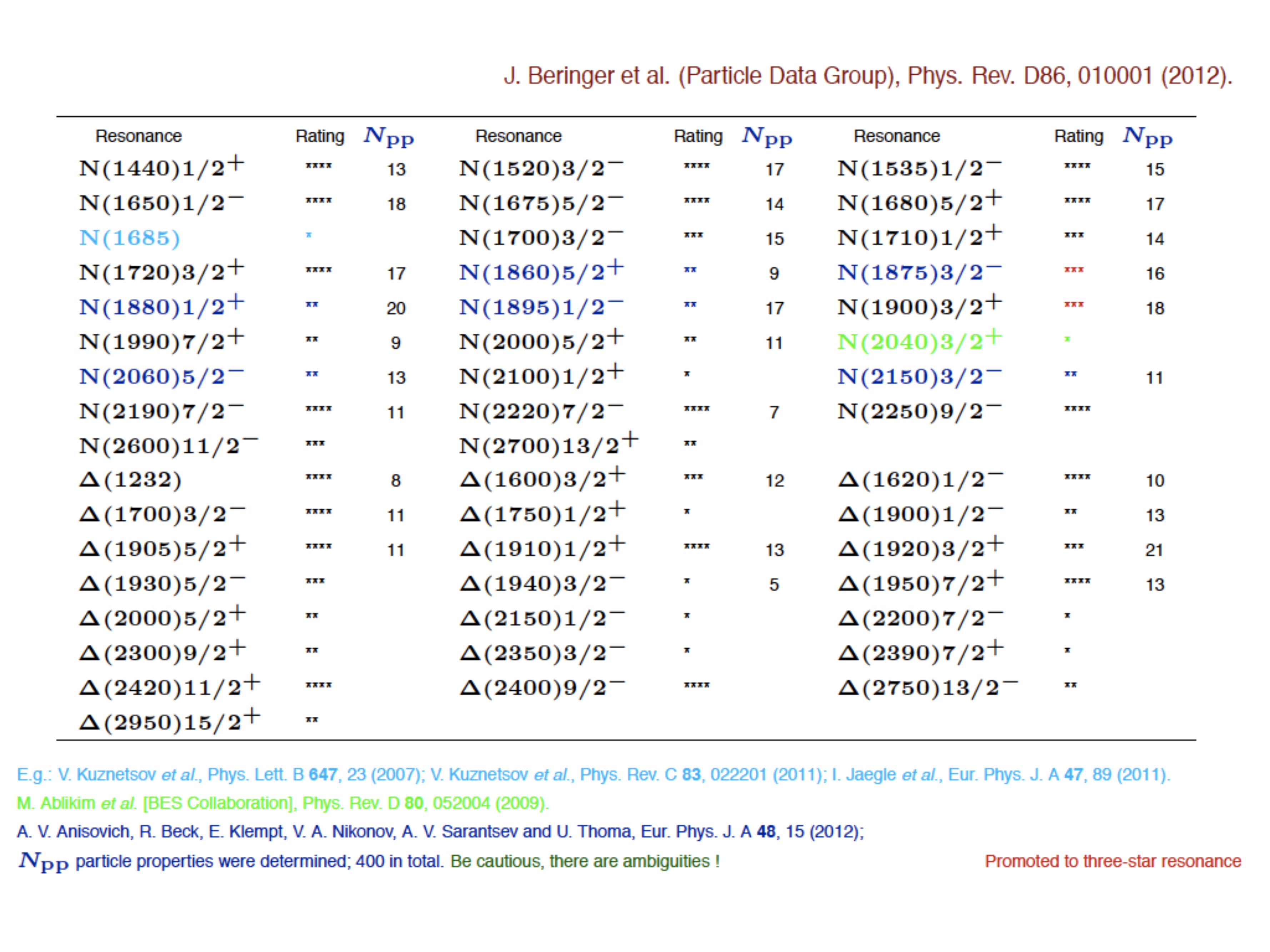} 
  \end{center} \vspace*{-10mm}
\caption{Baryon spectrum from the Particle Data Group with certain new states from the Bonn--Gatchina analysis~\cite{Anisovich:2012ct} and others. \label{fig:baryon}}
\end{figure}

The Bonn--Gatchina approach, formulated with covariant
amplitudes~\cite{Anisovich:2004zz}, performs combined analyses of all known data
on single and double-meson photo- and pion-induced reactions (see, e.g.,
Ref.~\cite{Gutz:2014wit}); four new states~\cite{Anisovich:2011fc} were reported
recently.  See Fig.~\ref{fig:baryon} for an overview.
In the Bonn--Gatchina approach, fits to reactions with two-body final states are
carried out by  minimization of $\chi^2$ functions, while the multi-body final
states  are analyzed in an event-by-event maximum likelihood method which  
fully takes into account all correlations in the multi-dimensional phase space.

The Gie{\ss}en group has recently included the analysis of $\pi\pi N$ data in form
of invariant mass projections~\cite{Shklyar:2014kra}, similar to the previous
work of the EBAC~\cite{Kamano:2008gr} group, while the original Kent
State~\cite{Manley:1984jz} analysis uses events directly.

In the search for excited baryons, considerable progress has been made in the
analysis of the corresponding data. In particular, recent data with
unprecedented accuracy from ELSA, JLab, MAMI, and other facilities have improved
the precision determination of resonance parameters. Still, no consensus has
been reached on the resonance content, in particular for broad resonances or
those that couple only weakly to the analyzed channels, cf.\ Fig.~\ref{fig:baryon}.
It is expected that additional constraints from crossed channels and analyticity
in complex angular momenta will help improve the reliability of resonance
extraction and determination of the spectrum. This is particularly relevant for
the data in forward direction and at higher energies. Here, a matching of Regge
amplitudes and unitary methods is a promising way to provide the correct
asymptotic behavior.

Another direction in which systematic uncertainties underlying these
phenomenological analyses can be quantified is to test whether the amplitudes
satisfy $S$-matrix analyticity as expressed by finite-energy sum rules (FESR).
Forward dispersion and other relations are included in the GWU/INS (SAID)
approach~\cite{Workman:1991xs}. 

Despite the rather involved phenomenology and the conceptual differences of the
discussed baryon analysis tools, there are indications that results become
eventually consistent among different groups~\cite{Anisovich:2013vpa}, and that
the long-sought determination of the baryon spectrum gets within reach. The
expected additional double polarization data from leading photoproduction
experiments should provide a more consistent picture of baryon states up to masses of about
2.2\,GeV.

\subsection{Duality and finite-energy sum rules} 
\vskip-5pt [G.~Fox, V.~Mathieu, A.~Szczepaniak] \vskip10pt
\label{sec:dual}

In the preceding sections we focused on those $S$-matrix properties
that are most important at low energies. Specifically we discussed
how, at the level of partial waves, to employ analyticity in order to
implement unitarity and use effective Lagrangians to implement various
symmetries.

\begin{figure}[t]
  \begin{center}
    \includegraphics[width=.8\linewidth]{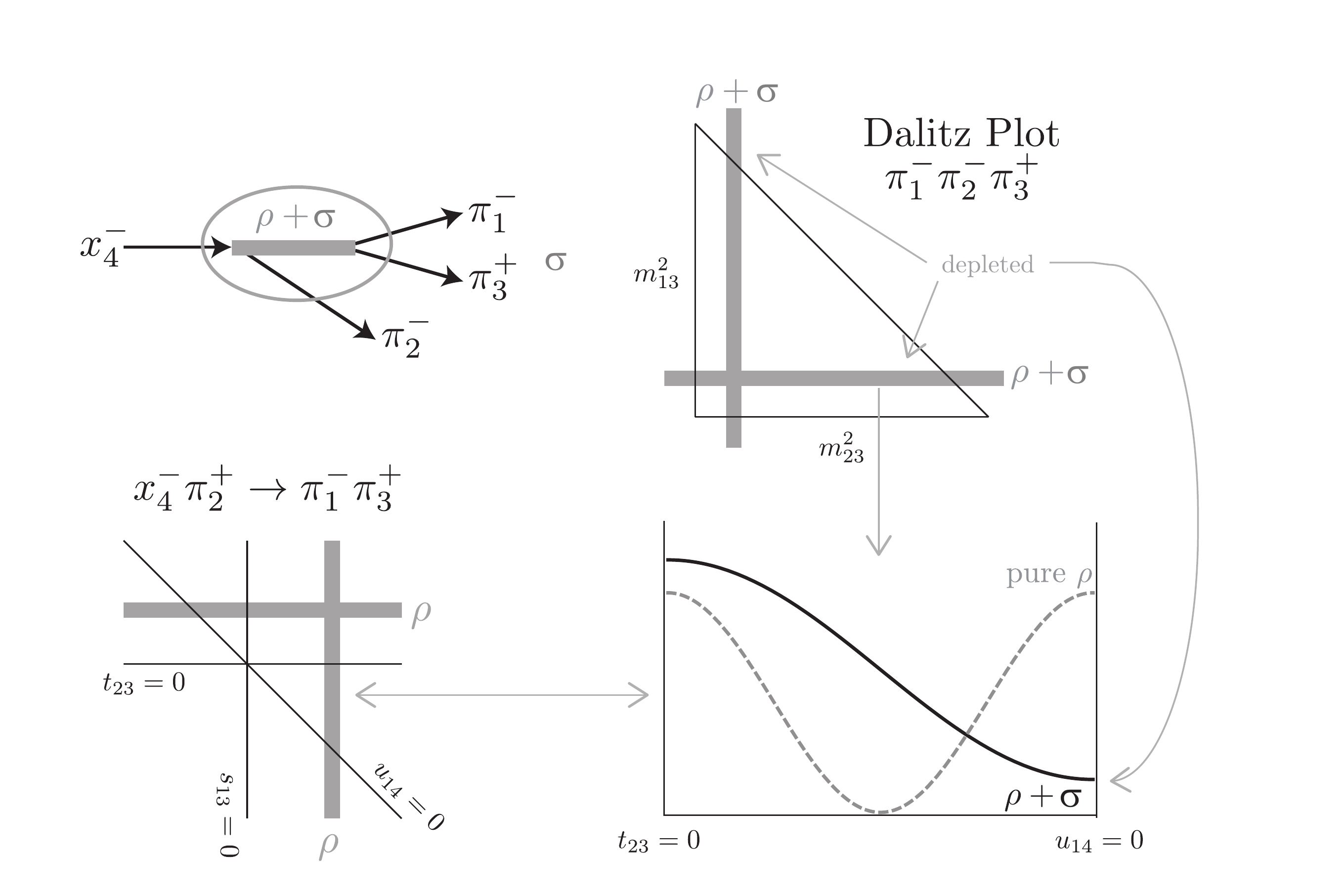} 
  \end{center} \vspace*{-3mm}
  \caption{The $\rho$ and  the $\sigma$ must interfere coherently to suppress
    double charge exchange in the $\pi^-_4 \pi_3^- \to \pi^-_1 \pi^-_2
    $ channel. The $\sigma$ refers to the lowest isospin-$0$ $\pi\pi$ resonance. \label{fig:EXD-GF}}
\end{figure}

The number of relevant partial waves grows with increasing channel
energy and in reactions that, at least in some channels, involve large
Mandelstam invariants a large (infinite) number of partial waves will
contribute. As shown by Regge, high-energy behavior in a direct
channel is dual to resonances in overlapping crossed channels. The
crossed-channel resonance contributions can be expressed in terms of
Regge poles and cuts, often referred to as Reggeons. The locations and
properties of Reggeons are constrained by analyticity of partial waves
continued to the complex angular momentum plane.
   
Schematically, as a function of channel energy variable, $s$, reaction
amplitudes can be separated into a contribution from the low-energy
region, where the $s$-dependence can be parameterized with a finite
number of partial waves, and the high-energy region, where the
amplitude is determined through Reggeons.  The low-energy partial
waves contain information about directly produced resonances, whereas
Reggeons know about resonances in crossed channels. To eliminate possible
double counting, the low-energy partial waves need to be removed from
the high-energy Reggeon contributions.  Analyticity is then used to
constrain the two regions. That is, with all other kinematical
variables fixed, the amplitude is an analytical function of channel
energy with singularities originating from bound states and opening of
physical thresholds. This enables one to write dispersion relations
that connect the low-energy partial waves with the high-energy
Reggeons.  The energy dependence of such DRs is often converted into a
set of moments and used as sum rules, also known as finite-energy sums
rules (FESR)~\cite{Collins:1971ff,Collins:1977jy} that relate
parameters of resonances in direct and crossed channels.  The classic
application of FESR was in charge exchange $\pi N$
scattering~\cite{Dolen:1967zz,Dolen:1967jr}, and used to establish a
relation between the leading, $\rho$~meson, and $\pi N$ resonances.

The observation that the low-energy contribution to FESR when
saturated by resonances reproduces the contribution from leading
Reggeons at high energy led to the concept of
duality~\cite{Mandula:1970wz,Phillips:1974mr}. According to this
hypothesis directly produced resonances in low partial waves are dual
to Reggeons, and residual, non-resonant backgrounds are dual to the
Pomeron.  This hypothesis is consistent with what is expected in the
limit of a large number of colors and the valence quark model.  It is therefore
worth noting that the existence of various exotic resonances that
cannot be accommodated within the quark model would also lead to
violations of this simple two-component duality. FESR studies can thus
provide additional arguments in favor of or against the existence of new
resonances. As an example let us consider $K^+p$ elastic scattering.
Directly produced resonances manifest themselves in the large
imaginary part of the amplitude. The $K^+p$ direct channel has
strangeness $+1$ and the absence of flavor exotic baryon resonances
implies relations between crossed-channel Reggeons that enforce the
vanishing of the Reggeon contributions to the imaginary part of the
amplitude.  These are known as exchange degeneracies (EXD), and in the
case of $K^+p$ involve the $\rho$ and $a_2$ Regge
trajectories. Similarly, destructive interference between the $\rho$ and the $\sigma$ (nowadays called $f_0(500)$) resonances in a direct channel is consistent with 
the absence of isospin $2$ resonances in $\pi\pi$ scattering in a crossed channel~\cite{RuizdeElvira:2010cs}.  The effect can be observed, for example, in the $3\pi$
Dalitz distribution obtained from $\pi^-$ diffractive dissociation, as
illustrated in Fig.~\ref{fig:EXD-GF}~\cite{GF}.

\begin{figure}
  \begin{center}
    \includegraphics[width=.4\linewidth]{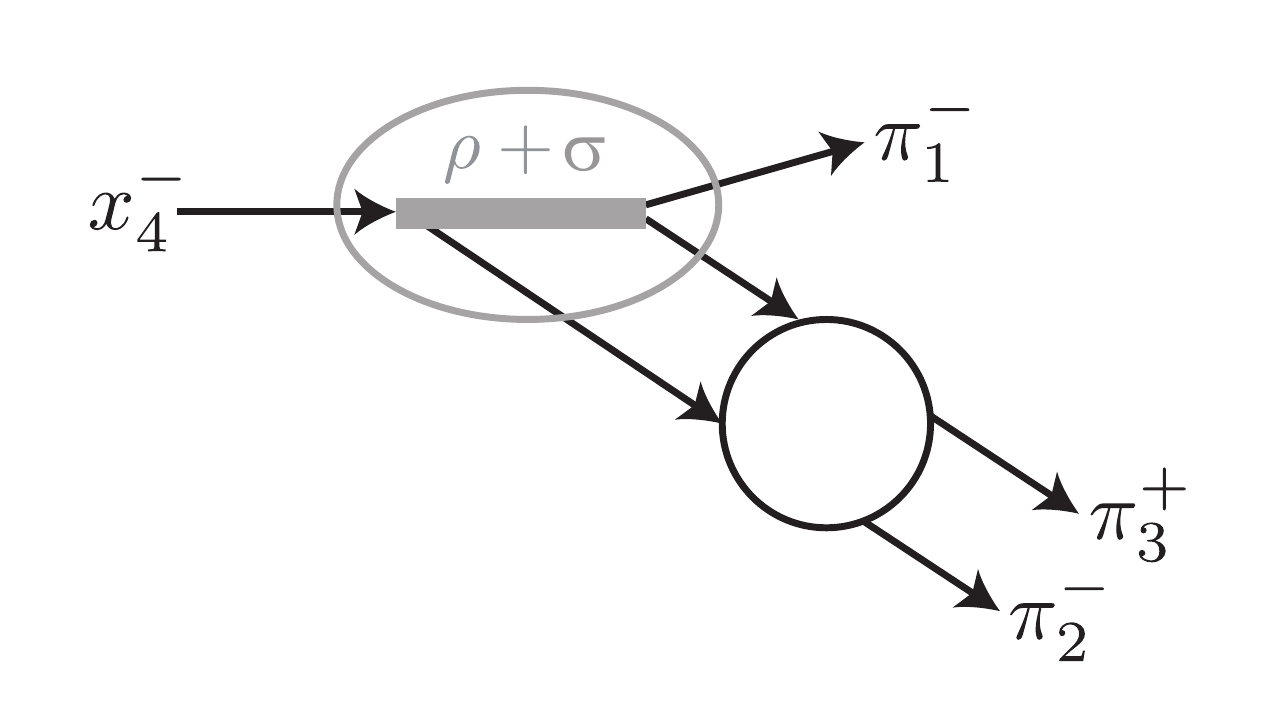}
  \end{center} \vspace*{-3mm}
\caption{This final-state interaction ``generated'' the Reggeons in the $23$ channel and we include these in the $\rho+\sigma$ ansatz in $13$~\cite{GF}. 
   \label{fig:FSI-GF} }
\end{figure}

FESR can also be use to distinguish what is background and what is
a $q\bar q$ resonance. As illustrated in Fig.~\ref{fig:FSI-GF} the
final-state interactions generating resonances in the low-spin partial
waves in the $23$ channel are dual to the $\rho^0$ and $\sigma$ in the $13$
channel.

\begin{figure}
  \begin{center}
    \includegraphics[width=0.35\linewidth]{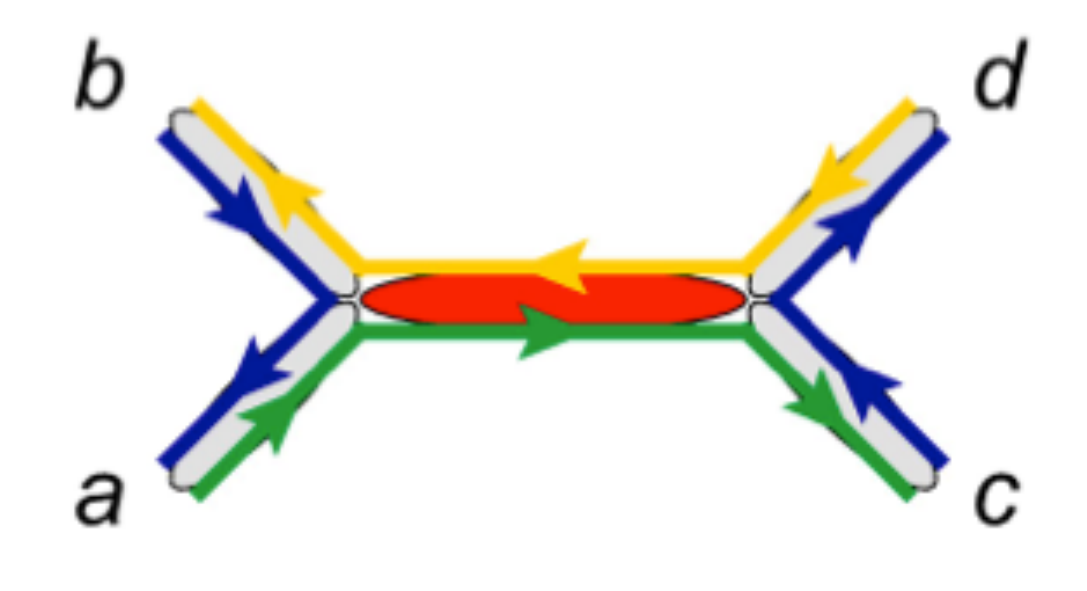}
    \includegraphics[width=0.3\linewidth]{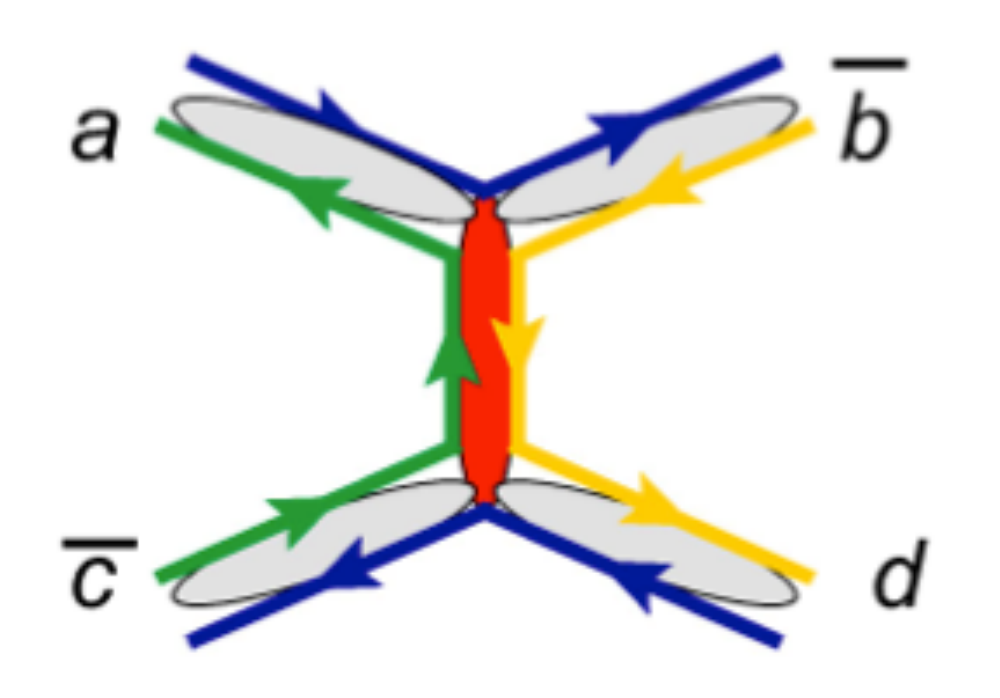}
    \includegraphics[width=0.3\linewidth]{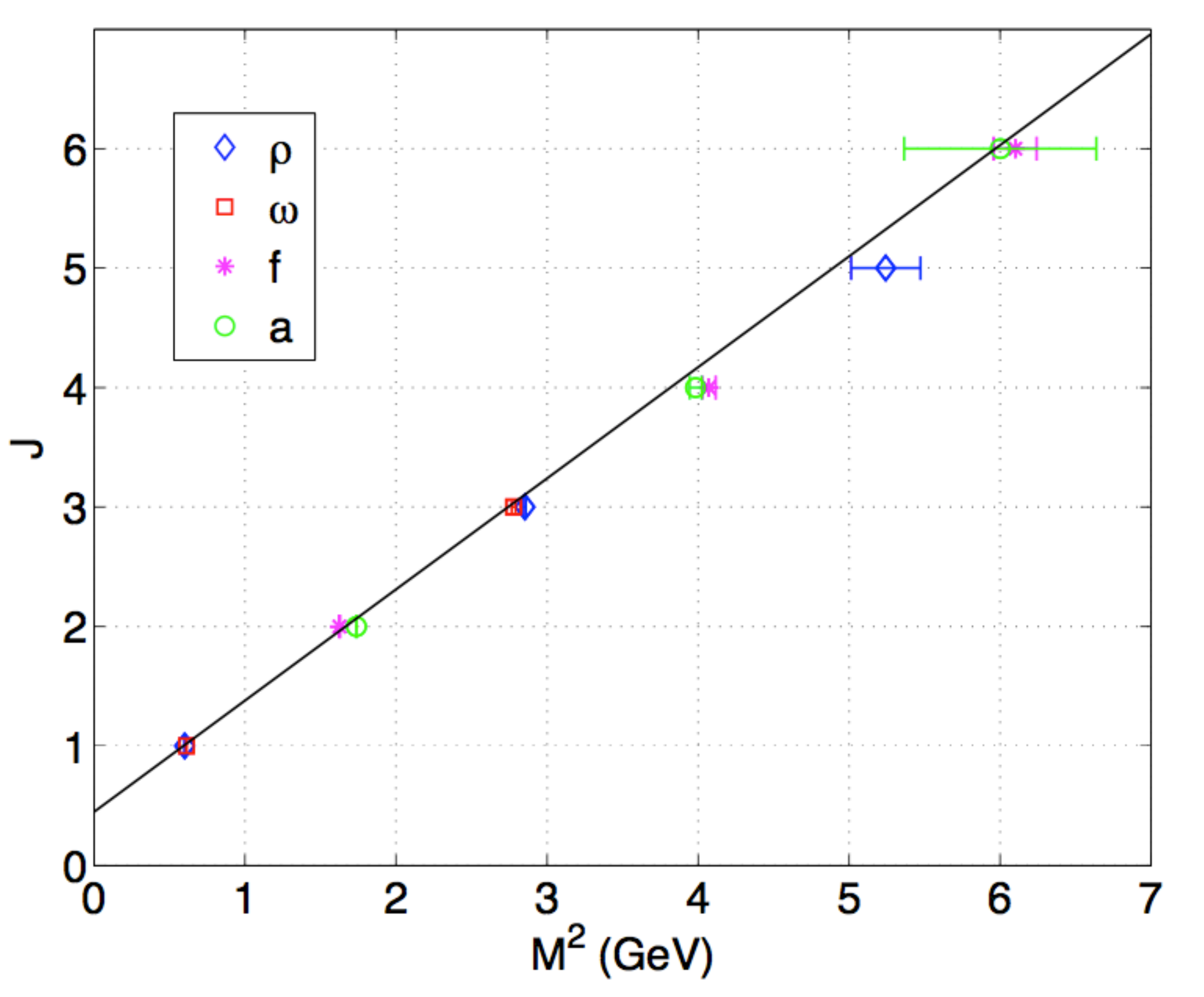}
  \end{center} \vspace*{-3mm}
  \caption{Duality hypothesis as supported by the quark model. The
    low-energy $s$-channel amplitude (left) is related to the
    high-energy $t$-channel amplitude (middle). Right: exchange
    degeneracy between the $\rho$, $\omega$, $f$, and $a$
    families. \label{fig:duality} }
\end{figure}

Exchange degeneracies between the leading Regge trajectories are
satisfied to within roughly $10\%$, and the EXD families are indicated
 in  Fig.~\ref{fig:duality}.  Duality therefore
leads to an important constrain that helps to reduce the number of
parameters in amplitude parameterizations and improve the
predictability of a fit.

\begin{figure}
\centering
  \includegraphics[width=0.35\linewidth]{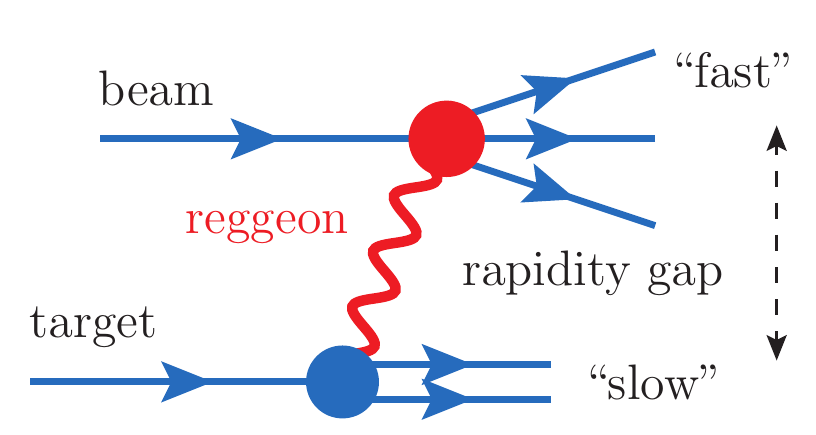}
\caption{Factorization in peripheral production. The upper vertex represents beam fragmentation 
and is described by beam plus Reggeon scattering. 
\label{fig:dp-1} }
\end{figure}

\begin{figure}
  \begin{center}
    \includegraphics[width=0.9\linewidth]{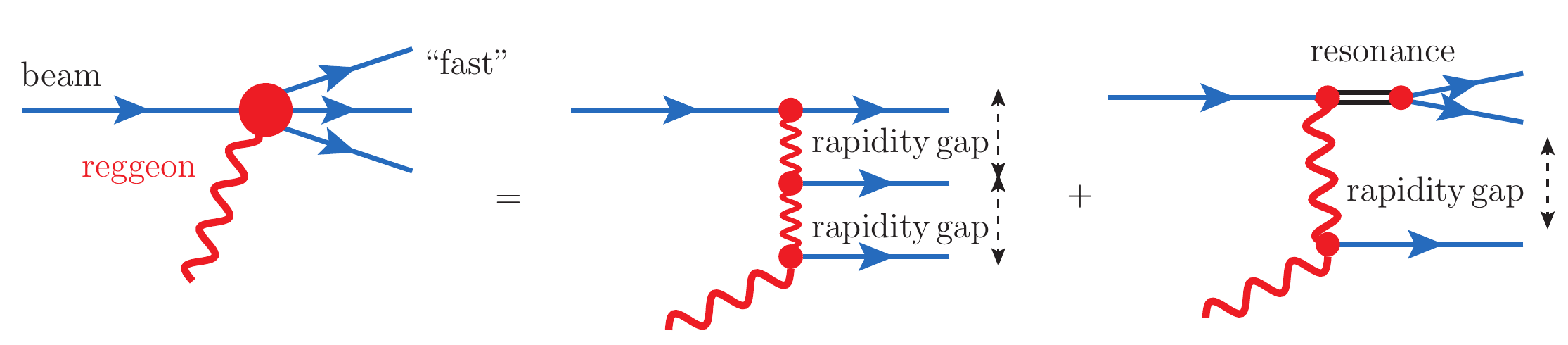}
  \end{center} \vspace*{-3mm}
\caption{Specific, non-overlapping contributions to the 
$\mbox{beam} \,+\, \mbox{Reggeon} \to 3\, \mbox{particles}$ amplitude. 
\label{fig:dp-2}}
\end{figure}

The resonance--Reggeon duality can be extended to multiparticle
production as illustrated in Figs.~\ref{fig:dp-1} and \ref{fig:dp-2}.  At
small scattering angle, when the center-of-mass energy of colliding
hadrons is significantly above the resonance region, the reaction
amplitude factorizes into a product of beam and target fragmentation
sub-processes mediated by the Pomeron/Reggeon exchange as depicted in
Fig.~\ref{fig:dp-1}.  With a meson or the photon as a beam and nucleon as
a target, beam fragments provide the laboratory to study meson
resonances, while the target fragments carry information about baryon
resonances.  Beam fragmentation has been the primary source of
information about meson--meson phase shifts (for instance from $\pi p\to \pi\pi p$~\cite{Grayer:1974cr} or from $K p\to K\pi p$~\cite{Aston:1987ir}) and two- and three-body
resonance decays. The description of the vertex representing
beam--Reggeon scattering to a few meson fragments follows the
principles of resonance--Reggeon duality. Again, Regge theory describes
interactions between hadrons at large values of relative energy and
angular momenta. It enables one to describe the bulk of the production
strength outside the resonance region. The latter is parameterized in
terms of a few partial waves at low masses and spins.  Parameters of
the low-spin partial waves can be fitted to data and self-consistency
between the low-energy (resonance) and high-energy (Regge) regions is
checked/enforced through finite-energy sum rules.

The leading Regge-pole dominance is an approximation, in principle
valid at asymptotically large channel energies. At finite energies,
the contribution from daughter trajectories and/or cuts may need to be
examined on a case-by-case basis.   
While the Regge amplitudes turn out to provide a good qualitative description
of the data even at energies as low as $2.5$\,GeV, a quantitative description
sets in only at significantly higher energies, cf.\ Ref.~\cite{Huang:2008nr} for
a detailed discussion of  $\pi N$ scattering---for a collection of 
earlier references see Ref.~\cite{Collins:1971ff}. Further research is
necessary to understand which scales control the energy/momentum transfer  ranges
where the Regge approach is working with the accuracy necessary to meet the goals 
outlined above. 
 
The cut contribution typically
accounts for diffraction in the final or initial state. To test the
Regge-exchange hypothesis one can also measure semi-inclusive
production: beam + target $\to$ leading particle(s) + $X$, where
``leading particles'' have large-$x$~\cite{Barnes:1978fw}. This is
described by some variant of Triple Regge coupling and so the
``leading'' part is beam + Reggeon $\to$ ``leading particles'' and so
similar to the case where $X$ is a simple particle such as
proton/neutron. Using arguments based on parton--hadron duality these
can be further related the partonic structure
functions~\cite{Bjorken:1973gc}.

\section{Tools} 
\vskip-5pt [R.~Mitchell, D.~G.~Ireland] \vskip10pt

One of the main challenges in experimental hadron spectroscopy is to
determine whether or not a given data set contains evidence to support
the existence of a previously unknown hadronic state (or states).  The
search for a new signal in experimental data consists of several
steps: defining a theoretical model of the data; maximizing a
likelihood function in order to fit the theoretical model to
experimental data; and performing statistical tests to evaluate how
well the model describes the data. The result of this process then
allows one to {\it decide} whether or not a new state has been found,
on the basis of model comparison.

While it is easy to list these steps, it has in the past not been
straightforward to carry out this analysis procedure without
incorporating approximations. The signals of interest are clearly not
large ones (otherwise they would have already been identified!), and
so we are at the stage of needing to move beyond crude approximate
methods. The three steps of constructing a likelihood function based
on a theoretical model, calculating the likelihood function with
measured data, and evaluating the goodness-of-fit all require a set of
tools that are both easy to use and contain state-of-the-art
methods. Each step presents challenges:

\begin{enumerate}
\item How to incorporate theoretical innovations into data models
  (likelihood functions)?
\item How to perform efficient calculations of likelihood functions?
\item How to use statistical methods to evaluate how well theory
  describes data?
\end{enumerate}

We now briefly summarize these issues, keeping in mind that the main
framework will be a partial-wave analysis (PWA) of experimental
data. In this, the key theoretical inputs are the {\it amplitudes} for
participating processes.  Afterwards, we list a few of the software
tools that are currently being used and ideas for future collaborative
code development.

\subsection{Incorporation of theoretical innovations}
\indent

The previous generation of amplitude analysis fitting tools had
several undesirable features: they commonly assumed the ``isobar
model'' with 2-body Breit--Wigner resonance decays; they were often
easy to use, but were also a sort of black box, offering little
flexibility to incorporate new amplitudes; and they were
model-dependent, where the model-dependence had unquantifiable
effects.

By contrast, the current generation of tools includes several
desirable features: they allow more flexibility when defining
amplitudes; they often force the user to explicitly code the
amplitudes, but are therefore less of a black box; they incorporate
state-of-the-art technology to increase fit speeds; and they allow
systematic studies of model dependencies.

There are no longer {\it experimental} or {\it technological} barriers
to incorporating theoretical innovations into experimental analyses.
Several software packages exist that can perform fits to experimental
data, using arbitrarily complicated amplitudes. An example of this is
the {\tt AmpTools}
package\footnote{http://sourceforge.net/projects/amptools} developed
at Indiana University, described further below.

\subsection{Efficient calculation of likelihood functions}
\indent

For statistical accuracy, the number of events that need to be
accumulated is ${\cal O} (10^6)$. In the search for a maximum of the
likelihood function, therefore, each change in the parameters of the
likelihood function will require ${\cal O} (10^6)$ evaluations of the
likelihood function. What is fortunate is that there are ways to make
use of the implicit parallelism in this calculation that utilize the
latest developments in hardware technology. The overall trend is from
multi-core to many-core processors, and from parallel to
massively-parallel computing.

The most promising avenue for PWA is general purpose graphical
processor unit (GPGPU) programming. Making use of the many cores on a
GPU, likelihood calculations can be performed on many chunks of data
at the same time. The pioneer approach of harnessing GPU parallel acceleration in PWA was performed in the framework of BESS-III~\cite{BESGPU}. Presently there are several hardware-specific
programming models (CUDA, OpenCL), but the field is in a state of
rapid development. Another potential game changer is Intel's Many
Integrated Core (MIC) architecture (Xeon Phi).

\subsection{Statistical evaluation of results}
\indent

Having obtained an unbinned maximum likelihood to obtain estimators
for any unknown parameters, the question is then ``How well does the
probability density function describe the data?'' Unfortunately,
an unbinned maximum likelihood does not provide any information that
would help answer this question. Typically we (somehow) determine the
``p-value." The p-value is the probability that a repeat of the
experiment would have lesser agreement with the data than what we
observe in our experiment.

In a binned analysis, this is often done by determining the $\chi^2$
statistic. In many analyses, though, binning is not a viable option
(due to high dimensions and/or low statistics). There are many methods
in the statistics literature that deal with these situations. However,
one must take care to choose the right tool for the job, and ensure
that one can properly validate any goodness-of-fit
test~\cite{Williams:2010vh}.

\subsection{Existing fitting tools and collaborative code development}
\indent

A number of software packages currently exist to aid in amplitude analysis fits.  Here we mention three: {\tt AmpTools}, {\tt ROOTPWA},
and {\tt MadGraph}.

{\tt AmpTools}, mentioned above, is a set of C++ classes that can be
used for amplitude analyses. The key class is the Amplitude class,
whose interface to the rest of the code is to take kinematics as input
and output a complex number. The user supplies as many of these as
needed. These amplitudes can be written either directly by theorists
or by experimentalists in collaboration with theorists.

\begin{sloppypar}
A new partial-wave analysis software package called {\tt
  ROOTPWA}\footnote{The software is available under GPL at
  http://sourceforge.net/projects/rootpwa/.} has been developed at TU~M\"unchen. 
The goal of this project is to provide a common package for
the analysis of multi-body final states produced in various reactions,
such as diffractive dissociation, central production, or
muo-production.  It includes a tool for the calculation of decay
amplitudes, which is an improved implementation of the helicity-based
isobar amplitude generator {\tt gamp} from the {\tt PWA2000} package
originally developed at BNL, augmented by scripts for automatic
symmetrization and testing. The amplitude calculator can be extended
to different spin formalisms and is in principle not limited to
isobar-like decay chains. The minimization is based on MINUIT2/MIGRAD
which comes as part of the {\tt ROOT} toolkit. {\tt ROOTPWA} is
completed by an $n$-body event generator and {\tt ROOT}-based
visualization tools.
\end{sloppypar}

For systematic amplitude generation, we can mention {\tt
  MadGraph}~\cite{Alwall:2011uj}. {\tt MadGraph}, developed at the
University of Illinois and at Louvain University, is a helicity
amplitude generator for tree-level Standard Model perturbation
theory. It is open source and easily modifiable to include effective
field theories.\footnote{http://madgraph.phys.ucl.ac.be} Events can be
generated with {\tt MadEvent}, and cross sections and other
observables can also be computed.

In order to make the best use of expertise to develop the best
open-source software, the programming community has over the years
evolved methods to make this collaboration work most efficiently. This
practice is gradually being taken up in the physics research community
as well. An outline of how a PWA community site might be structured is
as follows:

\begin{itemize}
\item Common code repository (can link to already existing sourceforge
  repositories) containing:
  \begin{itemize}
  \item Amplitude code
  \item Data readers
  \item Minimizers
  \item Integrators
  \item Plotters
  \item Parallelization libraries
  \item Exchange ideas (code snippets)
  \item Ecosystem of coexisting, independent codes
  \end{itemize}
\end{itemize}

\section{Concluding Remarks}

The new generation of experiments in hadron physics that are currently
under way or forthcoming will  continue
to generate complex data sets of very high quality (cf.\ Sec.~\ref{ch:experiments}).
Although quite involved analysis techniques are already now employed by
the various experimental groups (cf., e.g., Sec.~\ref{sec:anal-tech}), from the theoretical side an improved
understanding of the amplitudes used is
necessary  to analyze and interpret
the experimental results of hadron facilities in terms of resonance
parameters or, at higher energies, non-perturbative quark--gluon dynamics.
In addition
 a full command over hadronic interactions is
 important to hunt for physics beyond the
Standard Model. This
is especially obvious when it comes to CP~violation, which e.g.\ in the decays
of heavy mesons becomes visible only via the interference with
strong phases.

The key theoretical developments of the last decade were
in the realm of
\begin{itemize}
\item dispersion theory, especially for low- and medium-energy $\pi\pi$ interactions (up to $\sqrt{s}$ of about $1.4\,$GeV)
as well as pion rescattering in few-body final-state interactions (cf.\ Sec.~\ref{sec:disp}),
\item effective Lagrangian approaches, mainly for meson--baryon systems in
the resonance region  (cf.\ Sec.~\ref{sec:coup_chan}),
\item Regge theory, which is essential at high energies and for reactions in peripheral kinematics
 (cf.\ Sec.~\ref{sec:dual}).
\end{itemize}
Besides further improvements within the three approaches themselves, the
central goals for the future concern the merging of the different methods
and the reliable estimate of theoretical uncertainties.
For the former, it will be crucial to identify proper matching criteria in the kinematic 
regions of overlap, for the latter to find ways to systematically monitor
the accuracy of approximations made.

The aim of this document is to initiate or intensify  the discussion on
the methodology and tools needed to achieve these goals.  We expect
this discussion to continue through a series of workshops and schools
that are planned for the near future. We hope these will lead to the
development of state-of-the-art analysis tools that will become
available to practitioners of amplitude analysis techniques in the interpretation of
experimental data.

\end{document}